# A Dual Generalized Long Memory Modelling for Forecasting Electricity Spot Price: Neural Network and Wavelet Estimate

Souhir Ben Amor[1]*  Heni Boubaker[2]  Lotfi Belkacem[3]

**Abstract**

In this paper, a dual generalized long memory modeling has proposed to predict the electricity spot price. First, we focus on modeling the conditional mean of the series so we adopt a generalized fractional $k$-factor Gegenbauer process ($k$-factor GARMA). Secondly, the residual from the $k$-factor GARMA model has used as a proxy for the conditional variance; these residuals were predicted using two different approaches. In the first approach, a local linear wavelet neural network model (LLWNN) has developed to predict the conditional variance using two different learning algorithms, so we estimate the hybrid $k$-factor GARMA-LLWNN based back propagation (BP) algorithm and based particle swarm optimization (PSO) algorithm. In the second approach, the Gegenbauer generalized autoregressive conditional heteroscedasticity process (G-GARCH) has adopted, and the parameters of $k$-factor GARMA-G-GARCH model has estimated using the wavelet methodology based on the discrete wavelet packet transform (DWPT) approach. To illustrate the usefulness of our methodology, we carry out an empirical application using the hourly returns of electricity price from the Nord Pool market. The empirical results have showed that the $k$-factor GARMA-G-GARCH model has the best prediction accuracy in terms of forecasting criteria, and find that this is more appropriate for forecasts.

1 Institut des Hautes Etudes Commerciales de Sousse, Tunisia
*Correspondance author. Email: souhirbenamor@live.fr

2 IPAG LAB, IPAG Business School.

3 Institut des Hautes Etudes Commerciales de Sousse.

# 1 Introduction

In a modern society, electricity has become an essential commodity. Our daily lives depend on the use of electricity in various forms. Rapid rise of industrialization in the last century has contributed to a phenomenal growth of electricity consumption and therefore the tremendous increase in generation of electrical energy.

In power markets, price analysis has become an important topic for all its participants. Background information about the electricity price is crucial for risk management. More precisely, it represents an advantage for a market player facing competition. In fact, forecasting electricity prices at different periods is valuable for all industry stakeholders for cash flow analysis, financial procurement, capital budgeting, regulatory rule making, and integrated resource planning.

In this vein, both producers and consumers rely on price forecasting information to put forward their corresponding bidding strategies. If a producer has, an accurate price forecast, he can develop a bidding strategy in order to maximize his profit. On the other hand, if an accurate price forecast is available, a consumer can make a plan to minimize his own electricity cost. Hence, the players benefit is greatly affected by the accuracy of price forecast.

However, the behavior of electricity prices differs from other commodity markets. The most obvious of these differences is that electricity is a non-storable merchandize. Therefore, the demand and supply of electricity are highly inelastic and very sensitive to business cycles and weathers, and so relatively, small changes in load or generation in a matter of hours or minutes can cause huge changes in price.

The electricity spot prices exhibit also large and infrequent jumps caused by extreme load fluctuations (due to generation outages, severe weather conditions, transmission failures, etc.) [Clewlow and Strickland 2000; and Weron et al 2004]. It is exceptional that prices from one day to the next one, or even within just a few hours, can rise by a factor of ten or more. The period of considerable prices, are normally short, and prices tend to fall back down to "normal" levels after just a few hours. Such rapid price changes are of uttermost importance to take into consideration if one wants to understand and/or characterize the electricity spot price process. Moreover, electricity prices show some particular characteristics such as high frequency, multiple seasonality (on annual, weekly, and daily levels), non-stationary behavior, [Escribano et al 2011; Koopman et al 2007; and Knittel and Roberts 2005], high volatility, hard nonlinear behavior, long memory, calendar effect, high percentage of unusual prices, mean reversion, price spikes and limited information to the market participants. Hence, these behaviors may affect the prices dramatically. In this respect, there is no other similar market [Weron 2006]. Thus, we cannot rely on models developed for financial markets or other commodity markets. In this framework, due to the complexity of the electricity market, the electricity price forecasting has been the most challenging task. This has also motivated the researchers to develop intelligent and efficient approaches in order to forecast the prices thus that all stakeholders in the market can benefit out of it.

In this paper, we are interested in resolving the issues of modeling and forecasting the features of the electricity prices, notably, the existing of the seasonal long memory behavior in the conditional mean and the conditional variance.

In fact, most of the existing studies use models that permit the modelling of one or two features but not more. Specifically they do not model the long memory behavior inside the seasonality in both the conditional mean and the conditional variance. Recall that the long memory models introduced by [Granger and Joyeux 1981; and Hosking 1981] permit to model an infinite cycle, which is too restrictive for the electricity prices. One of the main characteristics of the high frequencies data sets is the presence of volatility clustering and leptokurtosis, as soon as persistence and cyclical patterns in the conditional mean of the series combined with conditional heteroscedasticity. All these characteristics have been presented inside electricity spot prices. Thus, dynamic modelling of means and variances appears essential for this kind of data sets. In this paper, we propose a new approach, which permit to take into account mainly all these features.

The novelty of our proposed method is its capacity to model the seasonal long memory behavior simultaneously in the conditional mean and in the conditional variance of electricity spot price, using a dual generalized long memory called $k$-factor GARMA-G-GARCH model. In addition, we adopt a wavelet estimation approach, which allows us to guarantee a parsimonious model with the greater accuracy. More precisely, the methodology of our dual generalized model consists into two steps; in the first step, the $k$-factor GARMA model, proposed by Woodward (1998), used to model the conditional mean in the time series. The choice of this model is motivated by the ability to take account simultaneously of long/short-term dependence and seasonal fluctuations at different frequency. The main feature of the $k$- factor GARMA model is that it allows larger diversity in the covariance structure of a variable witnessed by both the autocorrelation function and the spectral density function, which present $k$ singularities. By comparing it with the AFIRMA model, the $k$-factor GARMA model has the advantage of revealing certain aspects, by allowing periodic or quasi-periodic movements in the series.

Further, this approach focuses only on modeling the conditional mean, assuming that the residuals are white noises with a constant variance. However, it's well-recognized in the empirical studies that this hypothesis is not verified and the residuals are rather characterized by a time varying variance. To reproduce these patterns, we include in the second step the G-GARCH model, recently introduced by [Guégan 2000] in order to estimate the volatility, since this model is allows taking into account the long memory phenomenon and seasonality effect in conditional variance. Hence, we obtain the so-called $k$-factor GARMA-G-GARCH model. Then, for a comparison purpose, we use a second approach named the local linear wavelet neural network (LLWNN) model, proposed by [Chen, Yang and Dong 2006], as a nonlinear non-parametric method to estimate the conditional variance, so we obtain a hybrid $k$-factor GARMA-LLWNN model. More precisely, the $k$-factor GARMA model has shown its performance as a parametric model, with ability to capture all its information about the data within its parameters, the predicting data value from the current state of the model based only on its parameters. However, a non-parametric model can capture more subtle aspects of the data. It permits more information to pass from the current set of data attached to the model at the current state, to be able to predict any future data. The parameters have usually said to be infinite in dimensions and so can express the features in the data much better than parametric models. It has more degrees of freedom and is more flexible. Having observed more data will help to make an even better prediction about the future data. On the other hand, a parametric model has adopted as a parameter estimation method, when we have sufficient prior knowledge that the model has a parametric form with unknown parameters. While, the non-parametric models can reduce modeling bias by imposing no specific model structure other than certain smoothness assumptions, and therefore they are particularly useful when we have little information or we want to be re-flexible about the underlying model. To overcome the limitation of this parametric approach we combine a nonlinear non-parametric LLWNN model with the $k$-factor GARMA model to benefit the power of these two approaches in forecasting both conditional mean and conditional variance of electricity prices for the Nord Pool market. This paper provides three contributions. The main one consists in modeling the seasonal effect of electricity prices instead of using seasonal adjustment techniques and applying traditional prediction models in the deseasonalized time series. Therefore, the $k$-factor GARMA model has applied to analyze the seasonal long memory behavior. Secondly, we propose a dual generalized long memory model, so the $k$-factor GARMA model has extended to model the generalized long memory processes using a Gegenbauer GARCH model, we obtain the so-called $k$-factor GARMA-G-GARCH process, which allows for seasonal long-memory behavior associated with $k$-frequency. In fact, previous studies extend the $k$-factor GARMA model with the classic GARCH model or the FIGARCH model, and ignore the periodic long memory behavior when modeling the volatility. Concerning the parameters estimation of this model, we adopt an estimation method in the wavelet domain based on the maximal overlap discrete wavelet packet transform (DWPT) proposed by [Whitcher 2004].

Our final purpose is to show the performance of the proposed dual generalized long memory model (the $k$-factor GARMA-G-GARCH) by comparing it with the hybrid $k$-factor GARMA-LLWNN model, in order to prove the usefulness of modeling the seasonal long memory in both conditional mean and conditional variance to improve the forecasting accuracy of the electricity price. Moreover, in order to find the optimal architecture of the LLWNN we adopt two different learning algorithms, so we compare the back propagation (BP) algorithm and the particle swarm optimization (PSO) algorithm and then preserve the algorithm that minimizes errors.

The log-return of electricity price for the Nord Pool market is used in this paper in order to show the appropriateness and effectiveness of the proposed model to time series forecasting. The remaining part of the study is organized as follows; Section 2 presents a brief review of the literature. Section 3 present the econometric methodology; which includes the basic concepts of the $k$-factor GARMA model, the LLWNN model which is trained with two different learning algorithms; the BP and the PSO algorithms, and illustrate the methodologies describing the building of the hybrid $k$-factor GARMA-LLWNN model and the $k$-factor GARMA-G-GARCH model with the wavelet based estimation procedure. Section 4 deals with the empirical framework, where the proposed $k$-factor GARMA-G-GARCH model applied to log-return electricity price forecasting and its performance is compared with the hybrid $k$-factor GARMA-LLWNN model, and section 5 wrap up the conclusions.

## 2 Literature Reviews

The appropriate modelling of electricity price processes is of interest for several reasons. First, in electricity markets, the forecasting prices are interesting in the management and trading. Second, the operation of electricity markets has been considered similar to the operation of financial markets where electricity power derivatives are priced and traded in highly competitive markets. Thus, the modelling of both means and variances seems essential for this kind of data sets. Several methods were already been proposed in the literature to model them. In this vein, most of empirical works on electricity prices tend to focus on several features: high volatility, mean-reversion, spikes, and long-memory persistence.

An appropriate forecast model for electricity prices should consider the features described above. Meanwhile, important methods have been developed and applied for electricity price forecasting from different modeling families. The comparison of these different categories of models is significant, only if they were applied for the same analysis or problem. Otherwise, for a specific application, each model can show its strengths or weaknesses. In the field of electricity price forecasting, broadly two methods: the first approach is the statistical or econometric time series models, which considered as parametric tools, and the second approach is the soft computing models, which are considered as non-parametric tools. These two approaches are found to have been applied.

In statistical models, Auto regressive integrated moving average ARIMA [Contreras et al 2003], has been used extensively. However, these models do not allow taking into account the long memory behavior characterizing the electricity prices. To overcome this limitation Granger and Joyeux (1981), and Hosking (1981) introduced the Fractional Autoregressive Moving Average (FARMA) model. Recent works have applied these methods for the electricity prices [Koopman et al 2007; and Saâdaoui et al 2012]. In the frequency domain, these models present a peak for very low frequencies near the zero frequency. Hence, it is noteworthy that ARFIMA processes do not allow to take into account the cyclical behavior or persistent periodic in the data. To overcome this insufficiency, Gray et al (1989) introduced a second generation of the long-memory model denoted a generalized long-memory or Gegenbauer Autoregressive Moving Average (GARMA) model that has been developed to model, simultaneously, the presence of persistence and seasonality in the data. Such a model displays a hyperbolic decay of the autocorrelation function at seasonal lags rather than the slow linear decay characterizing the seasonal differencing model (SARIMA). More precisely, this model allows for a damped sinusoidal pattern in the autocorrelation function that decays toward zero at a hyperbolic rate. On the other hand, in the frequency domain, this model displays a spectral density that is not necessarily unbounded at the origin, as in the case of the ARFIMA model, but anywhere at a given frequency $\lambda$ along the interval $[0, \pi]$, $0 \leq \lambda \leq \pi$. This frequency named Gegenbauer frequency or G-frequency. The GARMA model exhibits a long-memory periodic behavior at only one frequency $\lambda$, thus implying just one persistent cyclical component. Recently, Woodward et al (1998) generalizes the single frequency GARMA model to the so-called $k$-factor GARMA model that allows the spectral density function to be not necessarily located at one frequency but associated with a finite number $k$ of frequencies in $[0, \pi]$, known as the Gegenbauer frequencies or G-frequencies. The main feature of this $k$-factor GARMA model is that it allows for more diversity in the covariance structure of a variable witnessed through both the autocorrelation function and the spectral density function, which presents $k$ singularities. Compared to the ARFIMA model, the $k$-factor GARMA model has the advantage of relaxing certain aspects, by allowing for periodic or quasi-periodic movements in the series. The $k$-factor GARMA model applied by several authors to reproduce the seasonal patterns as well as the persistent effects in the stock markets [Boubaker and Sghaier 2015; Caporale and Gil-Alana 2014; and Caporale et al 2012]. In addition, Ferrara and Guégan (2001) investigated the predictive ability of the $k$-factor Gegenbauer model on real data of Urban transport traffic in Paris area. Despite the compatibility of this model with the characteristics of electricity prices, few applications are oriented in the electricity market [Diongue et al 2009; Soares and Souza 2006; and Diongue, Dominique and Bertrand 2004].

However, the $k$-factor GARMA model exhibits two main shortcomings. The first is that it does not handle the nonlinear deterministic trend. Through simulation experiments, Beran (1999) proves that omitting the deterministic trend in the fractional integration model leads to a serious bias and high variability in the fractional integration parameter. The second is that it focuses only on modeling the conditional mean, assuming hence that the residuals are white noise with constant variance over time. In practice, this latter assumption is not verified and the residuals are rather characterized by a time-varying variance.

To reproduce these patterns, two approaches have been considered in the literature: the parametric models such as using GARCH or FIGARCH processes [Baillie et al 1996; and Bollerslev and Mikkelsen 1996], and the non-parametric methods like the neuronal nets for instance.

In the first category, Boubaker (2015) include the GARCH class of models proposed by Engle (1982) and Bollerslev (1986). Hence, the obtained model called the $k$-factor GARMA-GARCH process, which allows for long-memory behaviour associated with $k$-frequency and include a GARCH-type model to describe time varying volatility. In addition, Boubaker and Boutahar (2011) propose the $k$-factor GARMA-FIGARCH to reproduce the long-range dependence behaviour in the conditional variance of the exchange rate. Recently, Boubaker and Sghaier (2015) proposed a new class of semiparametric generalized long-memory models with FIAPARCH errors that extends the $k$-factor GARMA model to include nonlinear deterministic trend and allows for time-varying volatility, in some MENA stock markets, using a wavelet theory, estimation approach.

Nevertheless, these models are not fully satisfactory when modelling volatility of intra-daily financial returns series. One important characteristic of such data is the strong evidence of cyclical patterns in the volatility of the series. In fact, the periodic pattern appears as a persistent cyclical behaviour on the autocorrelation functions with oscillations decaying very slowly. Some pronounced peaks at one or more non-zero frequencies in the periodogram are also observed. The empirical evidence so far accumulated emphasises the importance of taking into consideration the periodic dynamics of volatility for a correct modelling. In order to model the empirical evidences of seasonal long memory behaviour in the volatility, Bordignon et al (2007, 2010) proposed new type of GARCH models characterised by periodic long memory behaviour. The suggested category of models introduces generalized periodic long-memory filters, which is based on Gegenbauer polynomials, into the equation of the standard GARCH model that describing the time-varying variance. These models, called periodic long-memory GARCH (PLM-GARCH), or generalized long memory GARCH (G-GARCH), and generalize the FIGARCH and FIEGARCH models, by introducing a reaches dynamics in the conditional variance. In fact, it accounts also for periodic long memory patterns in conditional variances (associated to the zero and non-zero frequencies of the power spectrum). In addition, to overcome highly non-linear coefficient constraints for variance positivity, the authors suggest modelling log-conditional variances. As a result, G-GARCH and PLM-GARCH nest some traditional long memory GARCH specifications when adapted to modelling log-variances. The filter used for G-GARCH specfications is the most general and allows the modelling of quite complex seasonal long memory behaviours. In the literature, the generalized long memory GARCH models (or G-GARCH) are used to estimate the financial time series such as the exchange rate by means of Monte Carlo simulations [Bordignon, Caporin and Lisi 2007; and Caporin and Lisi 2010]. Few studies apply this approach for the electricity spot price. To exemplify, Diongue and Guégan (2008) propose a new approach dealing with the stationary $k$-factor Gegenbauer process with Asymmetric Power GARCH noise under conditional Student-t distribution. This model called GGk-APARCH model is used to model electricity spot prices coming from some European and American electricity markets. With reference of forecasting criteria, this model shows very good results compared with models using heteroscedastic asymmetric errors.

We can conclude that in the literature of generalized long memory models, the authors use either the $k$-factor GARMA model, or the G-GARCH model to estimate the conditional mean and the conditional variance of the time series, respectively. However, none of them considers the existing of the long rang periodic behaviour in both the conditional mean and the conditional variance. In this paper, in order to provide robust forecasts for spot electricity prices, we propose a new approach based on dual generalized long memory process, which allows taking into account many stylized facts observed on the electricity spot prices, in particular stochastic volatility, long memory and periodic behaviours.

Concerning the estimation of the parameter's $k$-frequency GARMA process, Gray et al (1989) and Woodward et al (1998) considered the time-domain maximum likelihood method. On the other hand, Whitcher (2004) proposed an alternative estimation method in the wavelet domain based on the maximal overlap discrete wavelet packet transform (DWPT). Compared to Fourier analysis, the strength of the wavelet approach lies in its ability to localize simultaneously a process in both time and frequency.

At high scales, the wavelet has a small-centralized time support allowing it to concentrate on phenomena whish characterized by a short-lived time. On the other hand, at low scales, the wavelet has a large support enabling it to determine the long memory behavior. Hence, by moving from low scales to high scales, the wavelet zooms in on a process's behavior, identifying jumps, singularities and cups [Mallat and Zhang 1993; and Mallat 1999]. Consequently, the wavelet transform intelligently adapts itself to capture features across a wide range of time and frequencies. An important precondition for the construction of the wavelet-based estimator consists of the choice of the best-orthonormal-basis to diagonalize approximately the variance-covariance matrix of the $k$-factor GARMA process; Whitcher (2004) preconized the portmanteau tests at each level of wavelet decomposition. Boubaker (2015) evaluated the performance of a newly proposed estimation methodology for a $k$-frequency GARMA model in wavelet domain where the optimal orthonormal basis is selected by using the generalized variance portmanteau test to check for uncorrelated variance-covariance matrix.

In the second category, in order to grapple with the limitations of the parametric models and explain both the nonlinear patterns and time-varying variance that exist in real cases, several nonlinear, non-parametric models have been suggested. In this context, artificial neural networks have been extensively studied and used in modeling and forecasting electricity spot prices, Wang and Ramsay (1998) and Szkuta et al (1999), applied neural networks to model and forecast the dynamics of intra-day prices. The main advantage of neural networks is the ability to solve the complex nonlinear mapping problem, which possesses excellent robustness and error tolerance. Among the many existing tools, the ANN has received much attention because of its clear model, easy implementation and good performance in solving nonlinear problems, and this makes it suitable for modeling and forecasting of changing complex electricity system and electricity series. Hence, NNs have the ability to approximate any deterministic non-linear process, with little knowledge and no assumptions regarding the nature of the process. However, the classical sigmoid NNs have a series of drawbacks. Typically, the initial values of the NN's weights are randomly chosen; random weight initialization is generally accompanied with extended training times. In addition, when the transfer function is sigmoid function, there is always significant change that the training algorithm will converge to local minima. To overcome these limitations, another useful technique proposed in the recent years is wavelet based NN method. In this method, wavelet is merged with NN [Pati and Krishnaprasad 1993], and termed as wavelet neural network (WNN). Zhang and Benveniste (1992) developed the WNN, as an alternative to traditional NNs, in order to alleviate the abovementioned weaknesses. The WNN have one hidden layer networks which adopt a wavelet as an activation function, as an alternative of the conventional sigmoidal function. It is essential to note here that the multidimensional wavelets conserve the ''universal approximation'' property that characterizes NNs. Bernard, Mallat, and Slotine (1998) presented several reasons to prove the utility of adopting wavelets as a substitute of other transfer functions. Firstly, wavelets functions are characterized by high compression abilities, these are more appropriate for the modeling of high frequency signals. Secondly, computing the value at a single point or updating the function estimated from a new local measure, involves only a small subset of coefficients. Third, In contrast to the classical ''sigmoid NNs'', WNs allow for constructive procedures that efficiently initialize the parameters of the network. Using wavelet decomposition, a ''wavelet library'' can be constructed. In turn, each wavelon can be created by means of the best wavelet of this wavelet library. The main features of these procedures are: (i) the convergence to the global minimum of the cost function, (ii) the vector of initial weight is very close to the global minimum, and consequently drastically reduced training times, [Zhang & Benveniste 1992; and Zhang 1997]. Moreover, WNs offer information for each wavelon relative participation to the approximation function and the estimated dynamics of the generating process. Finally, an efficient initialization procedures will estimated the same weights vector which minimize the loss function for each time. The WNNs have been successfully applied in the field of time series prediction, [Cao et al 1995; and Cristea; Tuduce, and Cristea 2000] and in short-term load forecasting, [Bashir and El-Hawary 2000; Benaouda, Murtagh, Starck, and Renaud 2006; Gao and Tsoukalas 2001; Ulugammai, Venkatesh, Kannan, and Padhy 2007; and Yao, Song, Zhang, and Cheng, 2000]. However, the major drawback of the WNN is that it needs many hidden layer units for higher dimensional problems. In WNN theory, the curse of dimensionality is considered as an unsolved problem, thus it is difficult to apply the WNN to high dimensional problems. For that reason, the WNN are applied usually to problems with small input dimensions. This can be explained by the fact that they are composed of regularly translated and dilated wavelets. For the WNN, the number of wavelets drastically increases with the input dimension. In order to take advantage of the local capacity of the wavelet basis functions while not having too many hidden units, an alternative type of wavelet neural network known as local linear wavelet neural network (LLWNN) has been proposed by Chen and al (2004). This model replaces the connection weights between the hidden layer units and output units by a local linear model. Thus, it requires only smaller wavelets for a given problem than the case of wavelet neural networks. In addition, this local capacity offers certain advantages such as the efficiency and transparency of the learning structure. The LLWNN model has been used for the electricity price forecasting, [Pany 2011; Chakravarty et al 2012; and Pany and al 2013].

In fact, both the $k$-factor GARMA models, as a powerful statistical method, and the LLWNN model, as an advanced AI method, have achieved successes in their own nonlinear parametric and nonparametric domains respectively. However, none of them is a universal model that is suitable for all circumstances. In other words, a time series is often complex in nature and a single model may not be able to capture the different patterns in the same way, so no method is the greatest for all situations. Thus, using hybrid models or combining several models has become a common practice in order to overcome the limitations of model's components and improve the forecasting accuracy. In addition, since it is difficult to completely know the characteristics of the data in a real problem, hybrid methodology can be a good strategy for practical use especially when the combined models are very different. In general, the statistical methods help in dealing with voluminous datasets and neural network handle the non-linearity. In the literature, different combination techniques have been proposed in order to overcome the deficiencies of single models. The basic idea of the model combination in forecasting is to use each model's unique feature in order to capture different patterns in the data. The difference between these combination techniques can be described using terminology developed for the classification and neural network literature [Sharkey 2002]. Hybrid models can be homogeneous, such as using differently configured neural networks [Armano et al 2005; and Yu et al 2005]. It can be heterogeneous, with both linear and nonlinear models [Taskaya and Casey 2005]. In a competitive architecture, the objective is to represent different patterns in the time series by building an appropriate module that including different parts of the time series, and to be able to switch control to the most appropriate. For example, a time series may generally exhibit nonlinear behavior, however, according to the inputOconditions, this nonlinearity can change to linearity. On the threshold autoregressive models (TAR) two different linear AR processes are applied, each one change control among themselves according to the input values [Tong and Lim 1980]. An alternative is a mixture density model, also known as nonlinear-gated expert, which comprises neural networks integrated with a feedforward gating network [Taskaya and Casey 2005]. In a cooperative modular combination, the aim is to combine models to build a complete picture from a number of partial solutions [Sharkey 2002].

The hybrid techniques that decompose a time series into its linear and nonlinear form are one of the most popular hybrid models categories, which have been shown to be successful for single models. Zhang (2003) suggested a hybrid methodology that combines both auto-regressive integrated moving Average (ARIMA) and artificial neural network (ANN) models. The motivation of Zhang's hybrid model comes from the following perspectives. First, in practice it is difficult to determine if the time series is generated from a linear or nonlinear process; hence, this problem can be solved by combining linear and nonlinear models. Second, in fact, it is rarely that the time series are pure linear or nonlinear and generally incorporate both linear and nonlinear patterns, which a single model may not be sufficient to estimate such process. Therefore, this problem can be solved by combining both linear and nonlinear models. Third, several forecasting literature proved that no single method is the greatest for all situations, since the time series is often complex and an individual model may not be able to detect different features in the data. Consequently, the chance to detect different patterns in the data can be increased by combining different models. Khashei and Bijari (2010) affirmed that the motivation for combining such models derives from the assumption that an individual model may not be appropriate to estimate all the features in the time series. Valenzuela et al (2008) proposed to combine some intelligent methods such as fuzzy systems, ANNs, and evolutionary algorithms so that this hybrid model could outperform the forecast accuracy of those methods when applied separately. Tseng et al (2002) combined seasonal time series ARIMA model and feedforward neural network (FNN). Tan et al (2010) suggested a new price forecasting method based on wavelet transform combined with ARIMA and GARCH models applied on PJM and Spanish and electricity markets.

To sum up, in the literature, different combination techniques has been proposed to overcome the limitation of single model by means of hybrid models, which uses the strength of different methods in order to model the different feature existing in the data, and thus enhance the forecasting results. However, the mentioned hybrid methods combine models that are not able to capture some features of the electricity time series, such as the seasonal long memory behavior, non-linearity, etc. In order to grapple with the limitation of such models, we combine in this study the $k$-factor GARMA model with the LLWNN in order to use each model's unique strength and thus, to capture different patterns in the electricity time series.

## 3 Econometric methodologies

### 3.1 The $k$-factor GARMA model

The $k$-frequency GARMA model, proposed by Woodward et al (1998), generalizes the ARFIMA [Granger and Joyeux 1981; and Hosking 1981] model allowing periodic or quasi-periodic movement in the data. The multiple frequency GARMA model is defined as follows;

$$\Phi(L)\prod_{i=1}^{k}(I - 2v_{m,i}L + L^2)^{d_{m,i}}(y_t - \mu) = \Theta(L)\varepsilon_t , \qquad (1)$$

Where $\Phi(L)$ and $\Theta(L)$ are the polynomials of the delay operator $L$ such that all the roots of $\Phi(z)$ and $\Theta(z)$ lie outside the unit circle. The parameters $v_{m,i}$ provide information about periodic movement in the conditional mean ($m$), $\varepsilon_t$ is a white noise perturbation sequence with variance $\sigma_\varepsilon^2$, $k$ is a finite integer, $|v_{m,i}| < 1$, $i = 1,2,\ldots k$, $d_{m,i}$ are long memory parameters of the conditional mean indicating how slowly the autocorrelations are damped, $\mu$ is the mean of the process, $\lambda_{m,i} = \cos^{-1}(v_{m,i})$, $i = 1,2,\ldots k,$ denote the Gegenbauer frequencies ($G$-frequencies).

The GARMA model with $k$-frequency is stationary when $|v_{m,i}| < 1$, and $d_{m,i} < 1/2$ or when $|v_{m,i}| = 1$, and $d_{m,i} < 1/4$, the model exhibits a long memory when $d_{m,i} > 0$.

The main characteristic of model is given by the presence of the Gegenbauer polynomial

$$P_m(L) = \prod_{i=0}^{k}(I - 2v_{m,i}L - L^2)^{d_{m,j}} \qquad (2)$$

This polynomial maybe considered a generalized long-memory filter that models the long-memory periodic behavior at $k+1$ different frequencies. When thinking of the $\lambda_{m,i}$ as the driving frequencies of a cyclical pattern of length $S$, and $k+1 = [S/2]+1$, where $[.]$ stands for the integer part.

To highlight the contribution of $P(L)$ at frequencies $\lambda_m = 0$ and $\lambda_m = \pi$, equation (2) can be written as:

$$P_m(L) = (I-L)^{d_{m,0}} (I+L)^{d_{m,k}I(E)} \prod_{i=1}^{k+1}(I - 2v_{m,i}L - L^2)^{d_{m,j}} \quad (3)$$

Where $I(E) = 1$ if $S$ is even and zero otherwise and $k+1 = [S/2] + 1 - I(E)$.

For a GARMA model with a single frequency, when $v_m = 1$, the model is reduced to an ARFIMA (p, d, q) model, and when $v_m = 1$ and $d = 1/2$, the process is an ARIMA model. Finally, when $d_m = 0$, we get a stationary ARMA model.

Cheung (1993) determines the spectral density function and shows that for $d_m > 0$ the spectral density function has a pole at $\lambda_m = \cos^{-1}(v_m)$, which varies in the interval $[0, \pi]$ It is important to note that when $|v_m| < 1$, the spectral density function is bounded at the origin for GARMA processes, and thus does not suffer from many problems associated with ARFIMA models.

## 3.2 A local linear wavelet neural network (LLWNN) model

Chen, Yang and Dong (2006) propose a local linear wavelet neural network (LLWNN) model for time series forecasting, and they have shown that this model has more accuracy than the traditional WNN. In local linear wavelet neural network (LLWNN) the number of neurons in the hidden layer is equal to the number of inputs and the connection weights between the hidden layer units and output units are replaced by a local linear model.

Remind that in the wavelet transform theory the wavelets are in the following form

$$\varphi = \left\{ \varphi_i = |a_i|^{-1/2} \varphi(\frac{x-b_i}{a_i}) : a_i, b_i \in R^n, i \in Z \right\} \quad (4)$$

Where

$$\begin{aligned} x &= (x_1, x_2, \ldots, x_n), \\ a_i &= (a_{i1}, a_{i2}, \ldots a_{in}), \\ b_i &= (b_{i1}, b_{i2}, \ldots b_{in}). \end{aligned} \quad (5)$$

Are families of functions that have been generated by a unique function $\varphi(x)$ by the operations of dilation and translation of $\varphi(x)$; which is located in both time space and frequency space, called a mother wavelet and the parameters $a_i$ and $b_i$ are named scale and translation parameters, respectively. The set $x$ represents the inputs to the WNN model. In the standard form of WNN, the output of a WNN is given by:

$$f(x) = \sum_{i=1}^{M} \omega_i \varphi_i(x) = \sum_{i=1}^{M} \omega_i |a_i|^{-1/2} \varphi(\frac{x-b_i}{a_i}) \tag{6}$$

Where $\varphi_i$, is the activation function of wavelets of the $i^{th}$ unit of the hidden layer and $\omega_i$ is the weight connection of the $i^{th}$ unit of the hidden layer to the output layer unit. Note that for the $n$-dimensional input space, the basis function of the multivariate wavelet can be computed by the product of $n$ unique wavelet basis functions as follows:

$$\varphi(x) = \prod_{i=1}^{n} \varphi(x_i) \tag{7}$$

Obviously, the location of the $i^{th}$ units of the hidden layer has been determined by the scale parameter $a_i$ and the translation parameter $b_i$: According to previous research, the two parameters can either be predetermined by the basis of the Wavelet transformation theory or be determined by a learning algorithm. Note that WNN is a kind of neural network in the sense that wavelets represent the basic functions.

Note that an intrinsic characteristic of networks with a basic function is the local activation of the units of the hidden layer, so that the connection weights associated with the units can be considered as locally significant constant models whose validity for one input is indicated by activation functions. Compared with the multilayer perceptron neural network, this local capacity provides certain advantages such as the efficiency of the learning and the transparency of the structure. Due to the rigidity of the local approximation, many basis functions have to be taken to approximate a given system. One limitation of the WNN is that for problems of large dimensions many units of hidden layers are needed.

In order to take advantage of the local capacity of the basic wavelet functions while minimizing the number of hidden units, here we propose another type of WNN.

In the local linear wavelet neural network (LLWNN), the number of neurons in the hidden layer is equal to the number of inputs and the connection weights between the hidden layer units and the output units are replaced by a local linear model.

In the literature of LLWNN it is known that the local linear model provides a more parsimonious interpolation in a large dimension space and thus provides it with the ability of time series prediction. This local capacity of the LLWNN model offers some advantages such as the efficiency and transparency of the learning structure.

The architecture of the proposed LLWNN is shown in the figure 1, and its output from the output layer is given by:

$$y = \sum_{i=1}^{M} (\omega_{i0} + \omega_{i1}x_1 + \ldots + \omega_{in}x_n)\varphi_i(x)$$

$$= \sum_{i=1}^{M} (\omega_{i0} + \omega_{i1}x_1 + \ldots + \omega_{in}x_n)|a_i|^{-1/2} \varphi(\frac{x-b_i}{a_i}) \qquad (8)$$

Where $x = [x_1, x_2, \ldots, x_n]$,

Instead the simple weight $\omega_i$, a linear model has introduced

$$\upsilon_i = \omega_{i0} + \omega_{i1}x_1 + \ldots + \omega_{in}x_n \qquad (9)$$

The linear models $\upsilon_i (i=1,2,\ldots,M)$ are determined by the associated locally active wavelet functions $\varphi_i(x)$ ($i=1,2,\ldots,M$), thus $\upsilon_i$ is locally significant. The motivations of introducing local linear models in WNN are as follows: (1) local linear models were studied in some neuro-fuzzy systems and showed good performance, and (2) local linear models could provide a more parsimonious interpolation in large spaces when modeling samples are dispersed.

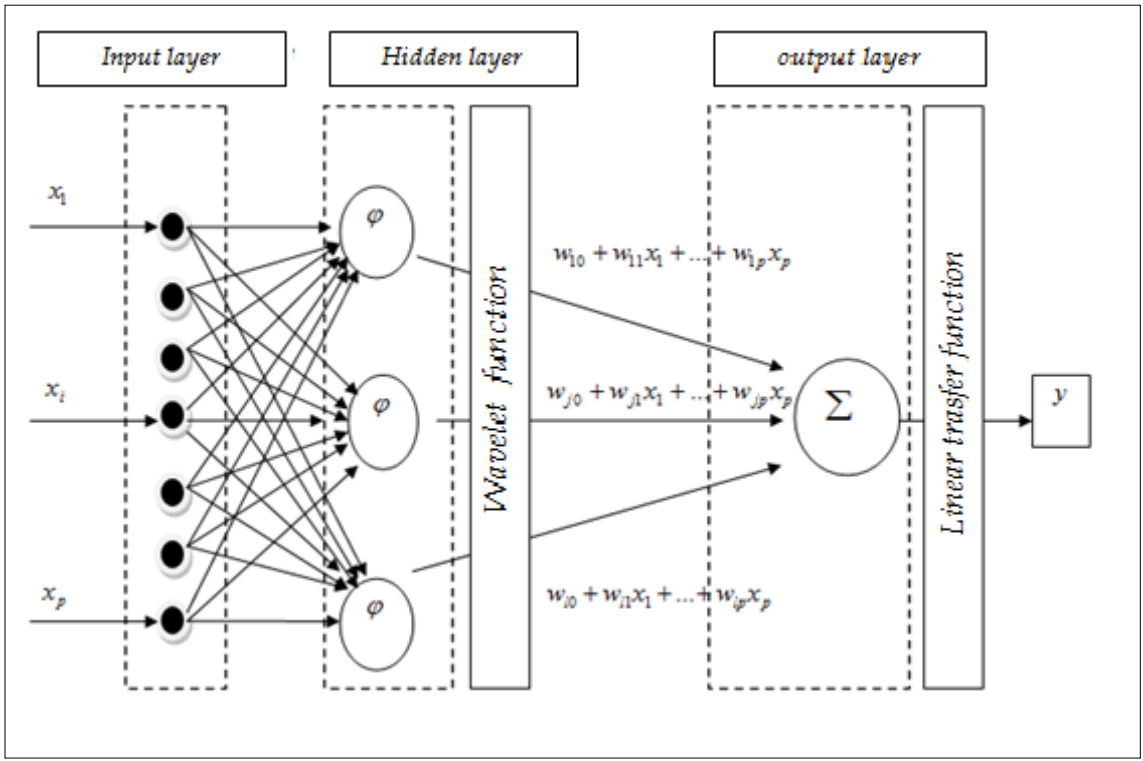

**Figure 1: Local Linear Neural Wavelet Network Architecture**

The scale and translation parameters of the local linear model are randomly initialized at the beginning and are optimized by a learning algorithm. In this paper, we adopt two different learning algorithms: the Back Propagation (BP) algorithm and the Particle Swarm Optimization algorithm (PSO), in order to preserve the best optimization method.

### 3.2.1. The back propagation learning algorithm for training the LLWNN model

The backward propagation of errors or back propagation, is classified as a supervised learning algorithm commonly used in training neural networks in general, in combination with an optimization method such as gradient descent.

The algorithm includes two iterative phase cycles; propagation and weight update. The input vector introduced to the network is propagated forward across the network, layer by layer, until it reaches the output layer. Afterward, the result of the output layer is compared to the target output, by means of a loss function. Hence, for each of the neurons in the output layer, an error value is calculated. This error represents the difference between the desired output and the outcome of the network. The error values are then propagated backward from the output layer until each neuron has an associated error value that approximates its contribution to the original output. These error values are used by the backpropagation to estimate the gradient of the loss function with respect to the weights in the network.

In the second step, this gradient is fed to the optimization method, which in turn exploits it to update the connection weights, in order to minimize the loss function. Briefly, the backpropagation algorithm minimizes the objective function by adjusting the connection weights employed to develop the models. The gradient of the cost function is calculated according to that particular weight parameter, and then the parameters are updated by means of the negative gradient.

The learning rate is a fundamental factor in the backpropagation algorithm. The network learns very slowly if the learning rate is too low, contrariwise, the weights and the objective function will diverge if this rate is too high. Therefore, an optimum value must be selected to guarantee global convergence, which tends to be a difficult task to achieve. If there are several local and global optima for the objective function, a variable learning rate will do better [Ham and Koslanic 2001]. Since the backpropagation requires a known and desired output for each value of input in order to determine the loss function gradient, for that reason it is usually considered as a supervised learning method.

The equations of the backpropagation algorithm are described briefly explained below. The objective function to minimize is given as:

$$E = \frac{1}{2}\left[y_t - \omega_{1,0}\varphi_1(x) - \omega_{1,1}p_1\varphi_1(x) - \ldots - \omega_{2,0}\varphi_2(x)\omega_{2,1}p_2\varphi_2(x) - \ldots \omega_{l,0}\varphi_l(x)\omega_{l,1}p_1\varphi_l(x) - \ldots \omega_{l,p}p_p\varphi_l(x)\right] \quad (10)$$

Where $y_t$ is the desired value, $\varphi(x)$ is the active wavelet function, $\omega_{1,0}$ denote the connection weight, $p$ is the number of input $(i=1,2,\ldots p)$ and $l$ is the number of the hidden units $(j=1,2,\ldots l)$. The weight is updated from $i^{th}$ to the $(i+1)^{th}$ iteration, that is from $\omega_t$ to $\omega_{t+1}$ is given by

$$\omega_{t+1} = \omega_t + \Delta\omega_t = \omega_t + \left(r\frac{\partial E_t}{\partial \omega_t}\right), \quad (11)$$

Denote that $r$ is the learning rate adopted in the LLWNN model.

Where $\dfrac{\partial E}{\partial \omega}$ for all weights are described by the following equations

$$\dfrac{\partial E}{\partial \omega_{i,0}} = \omega_{i,0} + r*e*\left(\dfrac{1}{2}\right)*(x_1^2 + x_2^2 + ... + x_p^2)*\exp(-((x_1-c_i)^2 + (x_2-c_i)^2 + ... + (x_p-c_i)^2)) \quad (12)$$

For $\forall j \neq 0$;

$$\dfrac{\partial E}{\partial \omega_{i,j}} = \omega_{i,j} + r*e*\left(\dfrac{1}{2}\right)*(x_1^2 + x_2^2 + ... + x_n^2)*\exp(-((x_1-c_i)^2 + (x_2-c_i)^2 + ... + (x_n-c_i)^2))*x_j \quad 13)$$

That is

$$\dfrac{\partial E}{\partial \omega_{1,0}} = \omega_{1,0} + r*e*\left(\dfrac{1}{2}\right)*(x_1^2 + x_2^2 + ... + x_p^2)*\exp(-((x_1-c_i)^2 + (x_2-c_i)^2 + ... + (x_p-c_i)^2)) \quad (14)$$

$$\dfrac{\partial E}{\partial \omega_{1,2}} = \omega_{1,2} + r*e*\left(\dfrac{1}{2}\right)*(x_1^2 + x_2^2 + ... + x_p^2)*\exp(-((x_1-c_i)^2 + (x_2-c_i)^2 + ... + (x_p-c_i)^2))*x_2 \quad (15)$$

$$\dfrac{\partial E}{\partial \omega_{2,0}} = \omega_{2,0} + r*e*\left(\dfrac{1}{2}\right)*(x_1^2 + x_2^2 + ... + x_p^2)*\exp(-((x_1-c_i)^2 + (x_2-c_i)^2 + ... + (x_p-c_i)^2)) \quad (16)$$

$$\dfrac{\partial E}{\partial \omega_{2,1}} = \omega_{2,1} + r*e*\left(\dfrac{1}{2}\right)*(x_1^2 + x_2^2 + ... + x_p^2)*\exp(-((x_1-c_i)^2 + (x_2-c_i)^2 + ... + (x_p-c_i)^2))*x_1 \quad (17)$$

The other weights are also updated in the same way (see figure 2).

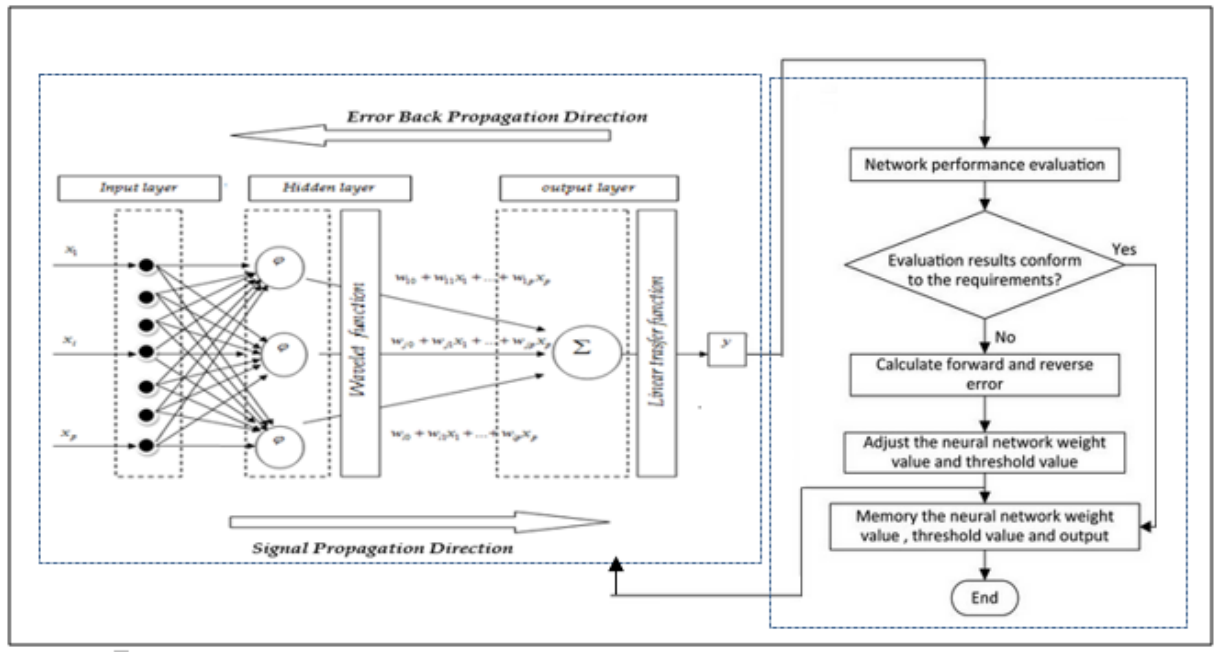

**Figure 2: Local Linear Wavelet Neural Network Model based Back Propagation Algorithm.**

### 3.2.2. The Particle Swarm optimization Algorithm (PSO)

Particle swarm optimization is developed through simulation of bird flocking in two-dimension space. The position of each agent is represented by $XY$ axis position and the velocity is expressed by $vx$ and $vy$. The agent position's modification is realized by the position and the velocity information.

The Bird flocking optimizes the objective function. Each agent knows its best value so far $(pbest)$ and its $XY$ position. Moreover, each agent knows the best value so far in the group $(gbest)$ among $(pbest)$. Mainly each agent tries to adjust its position using the following information.

    (a) The distance between current position and $pbest$.

    (b) The distance between the current position and $gbest$.

Velocity of each agent can be modified by the following equation:

$$v_i^{p+1} = wv_i^p + c_1 rand_1 \times (pbest_1 - s_i^p) + c_2 rand_2 (gbest - s_i^p) \tag{18}$$

Where $v_i^p$ is the velocity of agent $i$ at iteration $p$, $w$ is weighting function, $c_j$ is weighting factor, $s_i^p$ is the current position of agent $i$ at iteration $p$, $pbest_i$ is the $pbest$ of agent $i$ and $gbest$ is the $gbest$ of the group (see figure 3).

A certain velocity that progressively gets close to $pbest$ and $gbest$ can be calculated using the above equation. The current position, that represent the searching point in the solution space, can be adjusted using the following equation:

$$s_i^{p+1} = s^p + v_i^{p+1} \tag{19}$$

The first term of equation (18) is the previous velocity of the agent. The velocity of the agent is changed through the second and third terms.

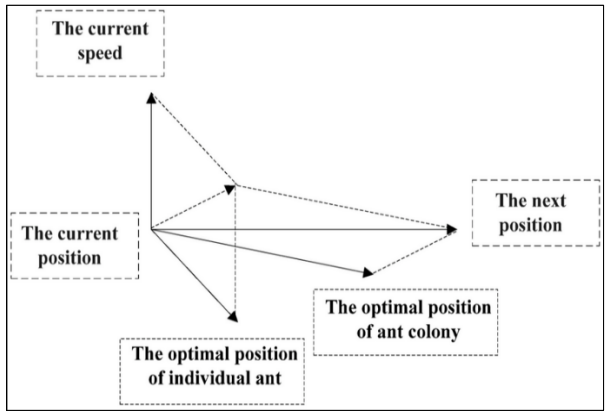

Figure 3: Particle Swarm Optimisation Algorithm.

The general steps that describing the optimization of the local linear wavelet neural network using the PSO algorithm can be illustrated as follows:

Step.1 For each agent the initial condition is generated:

The initial searching points ($s_i^0$) and velocity ($v_i^0$) of each agent are usually generated randomly within the allowable range. Note that the dimension of search space consists of all the parameters used in the local linear wavelet neural network as shown in equation (8).

The current searching point is set to *pbest* for each agent. The best-evaluated value of *pbest* is set to *gbest* and the agent number with the best value is stored.

Step.2 For each agent the searching points are evaluated:

For each agent the value of the objective function is calculated. If this calculated value is improved in comparison with the current *pbest* of the agent, the *pbest* value is replaced by the current value. If the best value of *pbest* is better than the current *gbest*, *gbest* is replaced by the best value and the agent number which present the best value is stored.

Step.3 Each searching point is modified:

Using equations (18) and (19), the current searching point of each agent is updated.

Step. The exit condition is verified:

If the number of the current iteration reaches the number of the predetermined maximum iteration, then exit;

If else; go to step 2 (see figure 4).

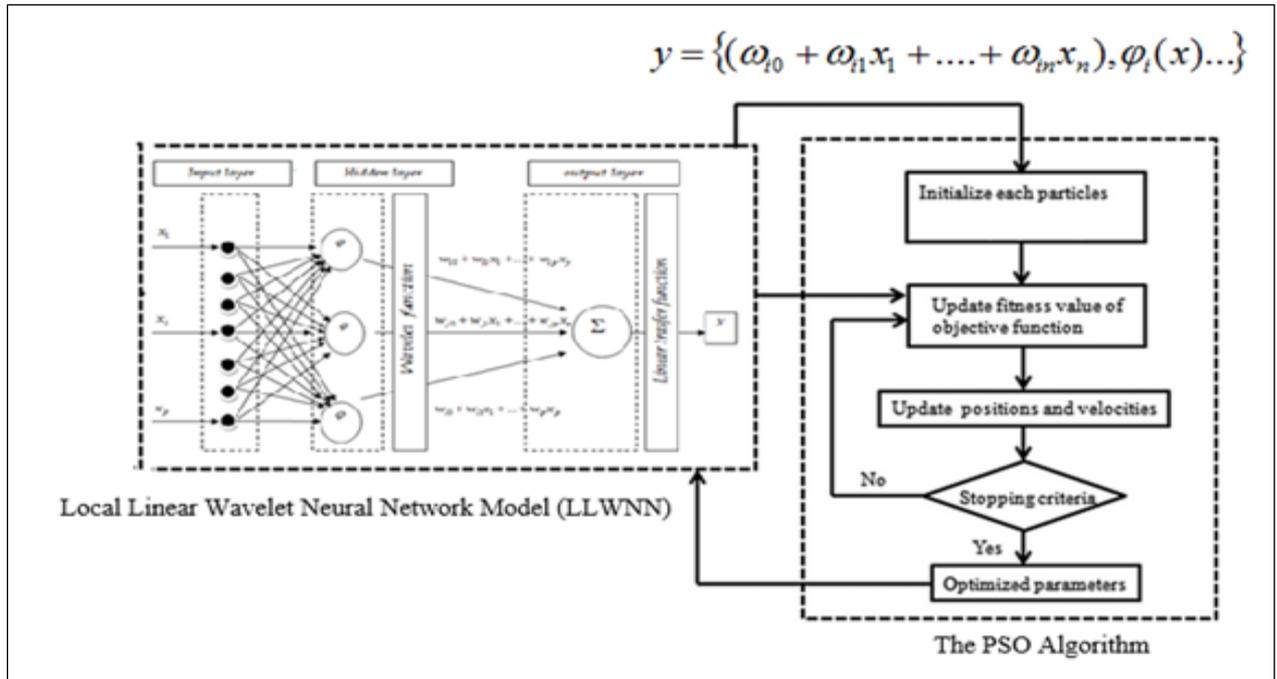

**Figure 4: Local Linear Wavelet Neural Network Model Based Particle Swarm Optimisation Algorithm.**

### 3.3  Hybrid *k*-factor GARMA-LLWNN model

Both theoretical and empirical results suggested that the combination of different models can be an effective tool to improve the predictive performance of each model, especially when the combined models are very different. In the literature, several combination techniques have been presented, such as the traditional hybrid model of Zhang (2003), the artificial neural network $(p,d,q)$ of Khashei et al (2010), and the generalized hybrid model of khashei et al (2011). These methods use the ARIMA model as a linear model and the multi-layer perceptron neural network to model the non-linear component. Although, these methods have shown, empirically, their effectiveness, insofar as they can be an effective way to improve the accuracy of predictions made by one of the methods used separately, these methods are critical.

Indeed, these methods exploited ARIMA modeling to predict the linear trend in the data, but predictions using the $ARIMA(p,d,q)$ model have not always proved to be very effective. The main criticism mainly concerns the modeling of short-term relationships only (short memory), while ignoring the seasonal effect and the presence of long memory that characterizes most financial and economic series.

To overcome this limitation, the $k$-factor GARMA model offers greater flexibility in modeling simultaneous short and long-term behavior of a seasonal time series.

On the other hand, the hybrid methods existing in the literature neglected the modeling of volatility, a phenomenon that characterizes most financial series. In reality, a good forecast must take into account the time varying variance. Thus, the LLWNN approach has been proposed to take into account the time varying of conditional variance. The choice of LLWNN in our hybrid model is motivated by the wavelet decomposition and its local linear modeling ability.

Furthermore, the previous hybrid models assume that the nonlinear relations are only exist in the residuals and the two components (linear and nonlinear) must be modeled separately, they assume that there are no nonlinear relations in the averages since they are always estimated using a linear model in the first step.

To overcome this limitation, we use the $k$-factor GARMA model to estimate the nonlinear components in the data, and then we model the residuals using an LLWNN model to predict the volatility. In other words, the first step consists in modeling the conditional mean using a non-linear parametric model (the $k$-factor GARMA). However, residuals are important in forecasting time series; they may contain some information that is able to improve forecasting performance. Thus, in the second step, the residuals resulting from the first step will be treated according to a local linear wavelet neural network LLWNN.

In our hybrid method, we have combined two models that have different characteristics in order to model the different characteristics existing in the data, thus we adopt a combination of parametric and nonparametric models.

In conclusion, the proposed hybrid model exploits the originality and the strength of the $k$-factor GARMA model as well as the LLWNN model to detect the different features existing in the data to benefit the complementary characteristics of the models, which compose them. Thus, the proposed hybrid model can be an efficient way giving a more general and accurate model than other hybrid models.

### 3.4 The $k$-factor GARMA-G-GARCH model

It is noteworthy, that the $k$-frequency GARMA model is considered as a generalized class of long memory models in that it relaxes several features of the ARFIMA model, by allowing for periodic or quasi-periodic movement in the series. Moreover, the $k$-factor GARMA model is stationary for a large class of processes and it is likely an enhanced alternative than seasonal differentiation. Nevertheless, this model assumes that the conditional variance is constant over time. In the empirical studies, it is well recognized that many time series often exhibit volatility clustering, where time series exhibit both high and low periods of volatility. To reproduce these patterns, we extended the $k$-factor GARMA model described above by inserting a fractional filter in the conditional variance equation. For this reason, we propose the $k$-factor GARMA-G-GARCH model that is able to capture both seasonality and long memory dependence in both the conditional mean and in the conditional variance.

In the last years, several studies discussed the phenomenon of long memory in the volatility of financial and economic time series. Several models were also suggested in the econometric and statistical literature to capture the persistence observed in the conditional variance; such as FIGARCH and FIEGARCH models [Baillie et al 1996; Bollerslev and Mikkelsen 1996; and Andersen and Bollerslev 1997] also the long memory stochastic volatility model [Breidt et al 1998], these models are well reputed and commonly adopted.

Recently, Bordignon et al (2007, 2010) introduced new GARCH-type models characterized by long memory and seasonality behavior, in order to model the empirical evidences of long memory behavior in the volatility of intra-daily financial returns. These models, named generalized long memory GARCH or Gegenbauer-GARCH (G-GARCH), generalize the FIGARCH and FIEGARCH models by introducing a seasonal long memory in the conditional variance.

In fact, it is also taken into account the periodic long memory patterns in the conditional variances, associated to the zero and non-zero frequencies of the power spectrum. Furthermore, to overcome the constraints of non-linear coefficients for the positivity of the variance, the authors propose to model the log-conditional variances. Therefore, G-GARCH nests some traditional specifications of long memory GARCH when adjusted to modeling log-variances.

The fundamental idea of this model is to include the generalized long-memory process into the equation describing the evolution of conditional variance in a GARCH framework. That's why this new class of models is called Gegenbauer-GARCH (G-GARCH). Thus, we consider the following $k$-factor GARMA process with G-GARCH type innovations to take into account the presence of a time varying conditional variance:

$$y_t = \mu_t + \varepsilon_t = \mu_t + \sigma_t z_t \tag{20}$$

Where $\mu_t$ is the conditional mean of $y_t$ modeling using the following $k$-factor GARMA model:

$$\Phi(L)\prod_{i=1}^{k}(I - 2v_{m,i}L + L^2)^{d_{m,i}}(y_t - \mu) = \Theta(L)\varepsilon_t \tag{21}$$

$$\varepsilon_t / I_{t-1} \sim N(0, \sigma_t^2) \tag{22}$$

Where $\sigma_t^2$ is the conditional variance, $I_{t-1}$ being the information up to time $t-1$, $z_t$ is an $i.i.d$ random variable with zero mean and unitary variance.

To specify the dynamics of the conditional variance, the starting point is the dynamics of $\varepsilon_t^2$. We assume that $\varepsilon_t^2$ follow a $k$-factor GARMA model, which describes a cyclical pattern of length $S$:

$$\left[(I-L)^{d_{v0}}(I+L)^{d_{v,k}I(E)}\prod_{i=1}^{k-1}(I - 2v_{v,i}L + L^2)^{d_{v,i}}\right]\psi(L)\varepsilon_t^2 = \gamma + [I - \beta(L)]\vartheta_t \tag{23}$$

$$P_v(L)\psi(L)\varepsilon_t^2 = \gamma + [I - \beta(L)]\vartheta_t \tag{24}$$

Where $\psi(L) = 1 - \sum_{i=1}^{q}\psi_i L^i$ and $\beta(L) = 1 - \sum_{i=1}^{p}\beta_i L^i$ are suitable polynomials in the lag operator $L$ and $\vartheta_t = \varepsilon_t^2 - \sigma_t^2$ is a martingale difference, $d_{v,0} = d_v/2$, $I(E) = 1$ if $S$ is even and zero otherwise.

With this assumption, the corresponding GARCH-type dynamics for conditional variance is given by;

$$\sigma_t^2 = \gamma + \beta(L)\sigma_t^2 + \left\{I - \beta(L) - \left[(I-L)^{d_{v,0}}(I+L)^{d_{v,k}I(E)}\prod_{i=1}^{k-1}(I - 2v_{v,i}L + L^2)^{d_{v,i}}\right]\psi(L)\right\}\varepsilon_t^2 \tag{25}$$

This implies that in the G-GARCH framework each frequency has been modeled by means of a specific long-memory parameter $d_{v,i}$ (differencing parameter of the conditional variance). When $d_{v,0} = d_{v,1} = \ldots = d_{v,k}$, all the involved frequencies have the same degree of memory. Model (25) may provide, in particular cases, most of the existing GARCH models. For example, standard GARCH models (included seasonal GARCH [Bollerslev and Hodrick 1992] can be obtained by putting $d_{v,i} = 0, \quad i = 0 \ldots k$. Similarly, the FIGARCH model is equivalent to $S = 1$ and $0 < d_{v,0} < 1$.

It is interesting to mention that generalized long-memory filters, in principle, may be applied to any category of GARCH structure. Nonetheless, due to the constraints needed for conditional variance positivity, G-GARCH models are not always feasible, for this reason, Bordignon et al (2007) proposed to model the logarithm of the conditional variances. Therefore, a practical computing solution is to apply the filter to a generalized log-GARCH model. This means beginning from the expression

$$P_v(L)\psi(L)\left[\ln(\varepsilon_t^2) - \tau\right] = \gamma + [I - \beta(L)]\vartheta_t \tag{26}$$

Where $P_v(L)$ is the generalized long memory filter introduced into a GARCH structure, $\vartheta_t = \ln(\varepsilon_t^2) - \tau - \ln(\sigma_t^2)$ is a martingale difference and $\tau = E\left[(\ln(z_t^2))\right]$. The expected $\tau$ value depends on the distribution of the idiosyncratic shock and ensures that $\vartheta_t$ is a martingale difference, given that $\ln(\varepsilon_t^2) = \ln(\sigma_t^2) + \ln(z_t^2)$. Under the Gaussian assumption $\tau = -1.27$.

The expression for conditional variance implied by (25) is:

$$\ln(\sigma_t^2) = \gamma + \beta(L)\ln(\sigma_t^2) + [I - \beta(L) - P_v(L)\psi(L)]\left[\ln(\varepsilon_t^2) - \tau\right] \tag{27}$$

Since we are modeling $\ln(\sigma_t^2)$ instead of $\sigma_t^2$, no constraints for variance positivity are necessary. A further approach of bypassing the problem of parameter constraints is to adopt EGARCH versions of our model.

The proposed model lies between the FIGARCH and the FIEGARCH representations. For the last model, the Stationarity of the covariance implies a memory parameter $0 \leq d_v < 0.5$ [Bollerslev and Mikkelsen 1996]. Differently, the covariance Stationarity of FIGARCH model is not obvious and is discussed in Giraitis et al (2000); Kazakeviéius and Leipus (2002); and Zaffaroni (2004), among others. The uncertainty on covariance stationarity existence extends to our model which, despite the dynamic formulation of the log-variance, is closer to the FIGARCH case rather than to the exponential model.

## 3.5 Wavelet-based estimation procedure

Concerning the estimation of the $k$-factor GARMA-G-GARCH model, we adopt an estimation procedure based on the wavelets following the methodology proposed by Whitcher (2004). Traditionally, the study of stochastic processes has been dominated by linear econometric methods that have proved useful, especially in finance. However, when dealing with hourly energy data, we cannot rely on models developed for financial markets or other commodity markets.

The complex behavior characterizes the spot price stimulated researchers to develop and test some methods for statistical analysis in order to realize a parsimonious model with greater accuracy. Recently, wavelet analysis is introduced as a very practical tool used to deal with complex phenomena. The ability of wavelets to capture variability over both time and scale can provide more insight into the nature of data on energy markets. These regimes can be effectively applied to decompose time series into different time scales. Hence, wavelet analysis is likely to reveal seasonality, discontinuities, and volatility aggregation [Gençay et al 2001]. For these reasons, wavelet transforms are applied for energy market data in order to capture nonlinear patterns and hidden patterns that exist between variables.

The advantage of this approach lies in its ability to simultaneously locate a process in both time and scale. At high frequencies, the wavelet has small centralized low scales, which allow it to focus on short-term phenomena. At low frequencies, the wavelet has a large time supports allowing it to identify long-term behavior. By moving from low to high frequencies, wavelets identify jumps and peaks [Mallat 1999; and Mallat and Zhang 1993].

The discrete wavelet packet transform (DWPT) [Whitcher 2004] is a generalization of the discrete wavelet transform DWT that splits the whole frequency band $[0, 1/2]$ into individual and regularly spaced intervals.

For a given temporal series $X$ of dyadic length $N = 2^J$, the $j^{th}$ level of DWPT is an orthonormal transform giving a vector of dimension $N$ of wavelet packet coefficients $(W_{j,2^{j-1}}, W_{j,2^{j-2}}, \ldots W_{j,0})'$ where each $W_{j,n}, n = 0, \ldots 2^j - 1$, has $N_j = N/2^j$ dimension and it is associated with the frequency interval $\kappa_{j,n} = [n/2^{j+1}, (n+1)/2^{j+1}]$.

Let $\{h_l\}_{l=0}^{L-1}$ and $\{g_l\}_{l=0}^{L-1}$ be the Daubechies wavelet and scaling filters respectively. Starting with the recursion $X = W_{0,0}$ were the $t^{th}$ elements of $W_{j,n}$ have calculated by the following steps of filtering

$$W_{j,n,t} = \sum u_{n,l} W_{j-1,[n/2],(2t+1-l) \mod N_{J-1_0}}, \quad t = 0, \ldots, N_j - 1 \tag{28}$$

Where

$$u_{n,l} = \begin{cases} g_l & \text{if } n \mod 4 = 0 \text{ or } 3 \\ h_l & \text{if } n \mod 4 = 1 \text{ or } 2 \end{cases} \tag{29}$$

Here [.] denotes the integer part operator.

It is notable that the collection of doublets $(j, n)$ (also called nodes) is known as a wavelet packet tree and will be denoted by $T = \{(j, n): j = 0 \ldots J; n = 0 \ldots 2^J - 1\}$. An orthonormal basis $B \subset T$ is obtained when a collection of DWPT coefficient vectors yield a disjoint and there is no overlapping complete covering of the frequency range $[0, 1/2]$ called a "disjoint dyadic decomposition".

Hence, in matrix notation, a vector of DWPT coefficients has obtained via $w_B = W_B X$ where $W_B$ an orthonormal $N \times N$ matrix is defining the DWPT through the basis $B$.

In this paper, to identify the "best base" $B$, from all possible orthonormal partitions, we use statistical white noise tests named the portmanteau test, following the method of Boubaker (2015). Similarly to the MODWPT, the step of the down sampling relative to the DWPT can be also removed by means of a variant of this transform identified as the MODWPT that depend on rescaled versions of the filter $u_{n,l}$ presented above.

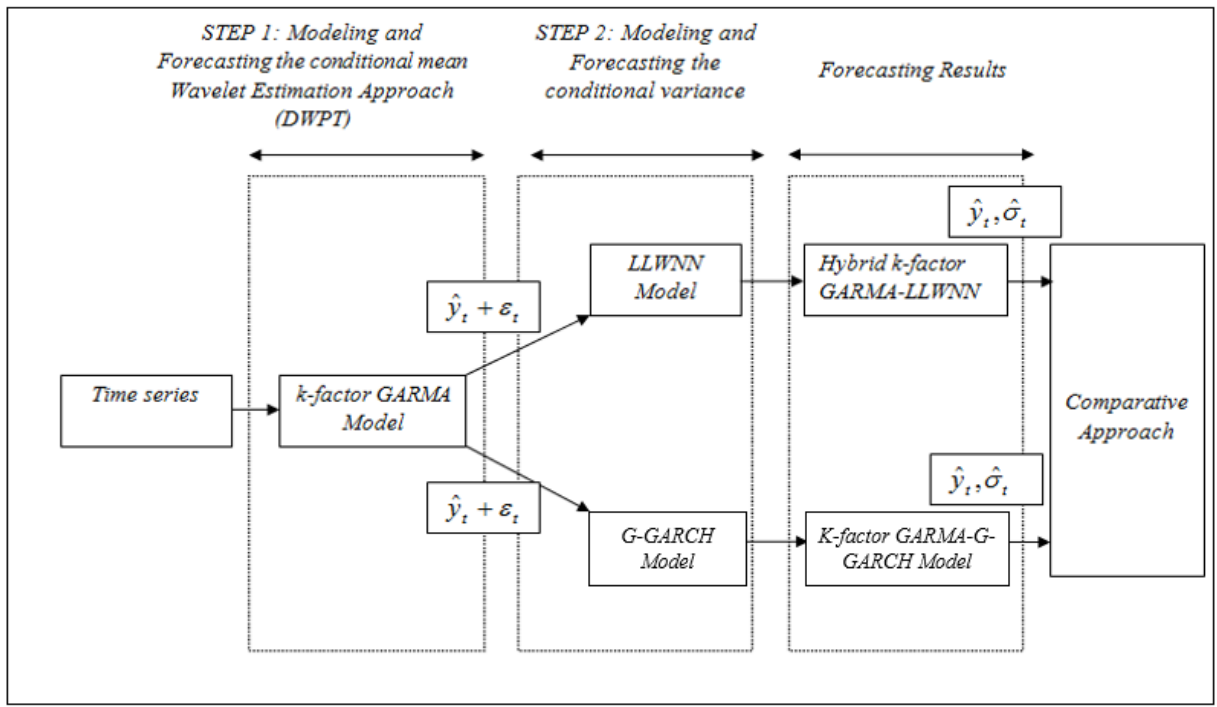

**Figure 5: A schematic representation of the adopted econometric methodology.**

## 4 Empirical results

### 4.1. The Nord Pool Electricity Market: Data description and Preliminary study

The Nordic electricity market known as Nord Pool market is a stock market affected to the electricity product. This stock market was created in 1992. It includes the three Scandinavian countries: Norway, Sweden and Denmark and Finland. In fact, this market began functioning officially in 1993 with Norway as the only area. Later, Sweden joined in January 1996. Finally, Finland was fully integrated in March 1999.

The spot price is the equilibrium price, and is calculated as the equilibrium point for every 24 hours. It is the unique price throughout the north region, and is determined when the supply and demand curves intersect. In case the data does not define an equilibrium point, the transactions will not take place.

In the electricity markets, a dynamic study of data is carried out without interruption. In fact, contrary to most other statistical studies in finance or economics, time series analysis is based on the assumption that the data represent consecutive measurements taken at equidistant time intervals 24h/24h. Although this assumption is violated for a large majority of financial data sets due to the market close, weekends and holidays, it is satisfied in electricity exchange market, mainly for spot prices, transaction volumes, production, etc. This allows statistical techniques to be applied appropriately and in agreement with the manner in which they were designed for use.

The methodology proposed in this research is tested on hourly return of spot prices on theNord Pool electricity market, covering the period between the 1st of January 2015 and 31st of December 2015, in total $N = 876$ hourly observations illustrated in figure 6. This data was extracted from the official website of Nord Pool market.

In this section, we analyze the spot price electricity series on the Nord Pool market, in order to study their statistical and econometric features. In most cases, the econometricians consider the logarithm of their series because the use of the difference logarithm sometimes makes the series stationary and allows modeling returns series. For this reason, we use the series of log-returns spot price (denote RSP).

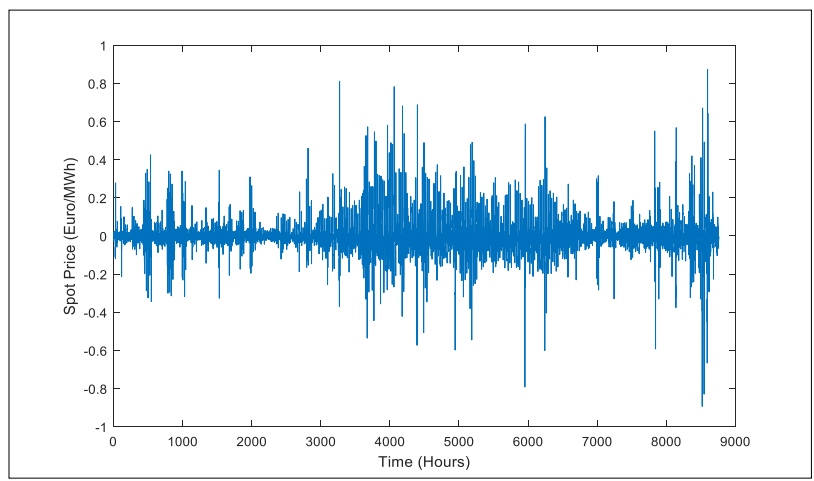

**Figure 6: Hourly spot price for NordPool electricity market.**

Figure 6 shows the time evolution of spot electricity price (RSP) indicates that this series seems stationary. This hypothesis can also be supported by the unit root tests (ADF, PP and KPSS). In addition, the series presents a clustering of volatility since periods of low volatility are followed by periods of high volatility. This is a sign of the presence of the ARCH effect in the series.

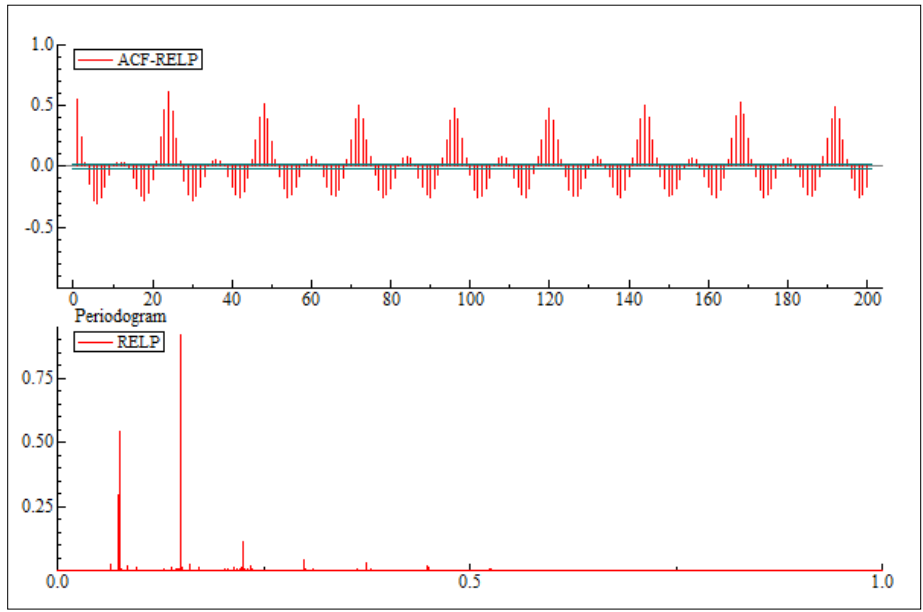

**Figure 7: Nord Pool ACF & Periodogram**

As shown in figure 7, for the return spot price electricity series, the spectral density, traced by the periodogram, shows peaks at equidistant frequencies, which proves the presence of several seasonalities in the conditional mean.

**Table 1: Descriptive statistics of the spot prices time series (log-returns).**

| | The NordPool log-returns |
|---|---|
| *Mean* | $-6.24 \cdot 10^{-5}$ |
| *Standard deviation* | 0.0835 |
| *Skewness* | 0.8838*** |
| | (0.000) |
| *Kurtosis* | 20.285*** |
| | (0.000) |
| *Jarque-Bera* | 110197 |
| | (0.000) *** |

*Note: levels of significance indicated between squared brackets. \*\*\* denotes significance at 1% level.*

The summary descriptive statistics of the Nord Pool log-returns are reported in table 1. The standard deviation is quite small, while the estimated measure of Skewness indicating a non-symmetric distribution. Moreover, the large value of the kurtosis statistic suggests that the underlying data are leptokurtic. This significant departure from normality is also confirmed by the large value of the Jarque-Bera (JB) test. Hence the electricity spot price series (RSP) is not normally distributed.

**The unit Root test**

**Table 2: ADF, PP, KPSS unit root testing results for Nord Pool log-returns.**

|  | *Model (3)* *With an intercept and a trend* | *Model (2)* *With an intercept* | *Model (1)* *Without an intercept* |
|---|---|---|---|
| *ADF Test* | -50.2201*** (0.000) | -50.2230*** (0.000) | -101.3755*** (0.000) |
| *PP Test* | -101.8959*** (0.000) | -101.8214*** (0.000) | - |
| *KPSS Test* | 0.0363 | 0.04104 | - |

*Note: *** indicate rejection of the null hypothesis at the 1-percent level.*
*ADF and PP: Critical values in the model (3): -3.95 (1%), -3.41 (5%), -3.12 (10%)*
*Critical values in the model (2): -3.43 (1%), -2.86 (5%), -2.56 (10%)*
*Critical values in the model (1): -2.56 (1%), -1.94 (5%), -1.62 (10%)*
*KPSS: Critical values in the model (3): 0.216 (1%), 0.146 (5%), 0.119 (10%)*
*Critical values in the model (2): -0.739 (1%), -0.463 (5%), -0.347 (10%)*

We tested for stationarity by performing unit root tests, namely, the augmented Dickey-Fuller (ADF), the Phillips-Perron (PP) and Kwiatkowski, Phillips, Schmidt, and Shin (KPSS) tests. These tests differ in the null hypothesis. The null hypothesis of the ADF and PP tests is that a time series contains a unit root, while the KPSS test has the null hypothesis of stationarity. The results of these tests are reported in table 2. The results of the ADF and PP unit root tests indicate that the Nord Pool log-returns time series is significant to reject the null hypothesis of non-stationarity. Thus, this series is stationary and suitable for subsequent tests in this study. Additionally, the statistics of the KPSS test support the acceptance of the null hypothesis of stationarity. Hence, the series is stationary, and suitable for long memory tests.

**The GPH (Geweke and Porter-Hudak, 1983) and LW (Robinson, 1995) long-range dependence**

Using the GPH [Geweke and Porter-Hudak 1983] and LW [Robinson, 1995] statistics, we test for the long-range dependence presence in the conditional mean. Corresponding results shown in Table 3 indicate evidence of long memory.

**Table 3: Results of GPH and LW long- range dependence tests in the conditional mean.**

|  | Bandwidth | GPH | | | LW | | |
|---|---|---|---|---|---|---|---|
|  |  | $\hat{d}$ | Standard error | p-value | $\hat{d}$ | Standard error | p-value |
| RSP $T=8760$ | $T^{0.5}=94$ | -0.2347 | 0.0718 | 0.001 | -0.236 | 0.0515 | 0.000 |
|  | $T^{0.6}=232$ | -0.3632 | 0.0440 | 0.000 | -0.417 | 0.0328 | 0.000 |
|  | $T^{0.7}=575$ | -0.3389 | 0.0273 | 0.000 | -0.328 | 0.0208 | 0.000 |
|  | $T^{0.8}=1425$ | -0.3810 | 0.0172 | 0.000 | -0.594 | 0.0132 | 0.000 |

**4.2. The estimation results**

**Table 4: Estimation of the $k$-factor GARMA model: a wavelet based approach.**

| Parameters | k-factor GARMA model estimation |
|---|---|
| $\hat{\Phi}$ | 0.0357*** (0.000) |
| $\hat{\Theta}$ | - |
| $\hat{d}_{m,1}$ | 0.2657*** (0.000) |
| $\hat{d}_{m,2}$ | 0.1238*** (0.000) |
| $\hat{d}_{m,3}$ | 0.0873*** (0.000) |
| $\hat{\lambda}_{m,1}$ | 0.1295*** (0.000) |
| $\hat{\lambda}_{m,2}$ | 0.0882 *** |
| | (0.000) |
| $\hat{\lambda}_{m,3}$ | 0.0479*** |
| | (0.000) |

*Note: *** indicate rejection of the null hypothesis at the 1-percent level.*

It's well known that electricity spot price often exhibits seasonal fluctuations (on the annual, weekly and daily level). The seasonal character of the prices is a direct consequence of demand fluctuations that mostly arise due to changing climate conditions like temperature or the number of daylight hours. Furthermore, the supply side shows seasonal variations in output. Therefore, these seasonal fluctuations in demand and supply translate into seasonal behavior of electricity prices, and spot prices in particular.

The seasonality can be easily observed in the frequency domain $\lambda_{m,i}=1/T$; where $\lambda_{m,i}$ is the frequency of the seasonality and T is the period of seasonality. As shown the spectral densities, represented by periodogram (figure 7), are unbounded at equidistant frequencies, which proves the presence of several seasonalities. They show special peaks at frequencies $\hat{\lambda}_{m,1}=0.1295$ (T= $7.72\approx 8$ hours $\approx 1/3$ day), $\hat{\lambda}_{m,2}=0.0882$ (T=11.5$\approx$12 hours $\approx 1/2$ day), and $\hat{\lambda}_{m,3}=0.0479$ (T=20.87 hours $\approx 1$ day), as shown in table 4, that corresponding to cycles with third daily, semi-daily and daily periods, respectively.

In the second step, the squared log-returns are used as a proxy of the corresponding volatility. Long memory tests are performed for the resulted time series. As reported in Table 5, the results of the GPH and LW indicate the presence of long memory in the conditional variance.

**Table 5: Results of GPH and LW long range dependence tests in the conditional variance.**

| | | GPH | | | LW | | |
|---|---|---|---|---|---|---|---|
| | Bandwidth | $\hat{\xi}$ | Standard error | p-value | $\hat{\xi}$ | Standard error | p-value |
| | $T^{0.5}$=94 | -0.0701 | 0.0718 | 0.3293 | -0.1164 | 0.0515 | 0.0240 |
| | $T^{0.6}$=232 | -0.9439 | 0.0440 | 0.0000 | -0.7501 | 0.0328 | 0.0000 |

| RSP | $T^{0.7}=575$ | -1.0848 | 0.02737 | 0.0000 | -0.7834 | 0.02085 | 0.0000 |
|---|---|---|---|---|---|---|---|
| $T=8760$ | $T^{0.8}=1425$ | 0.5481 | 0.01723 | 0.0000 | 1.4457 | 0.01324 | 0.0000 |

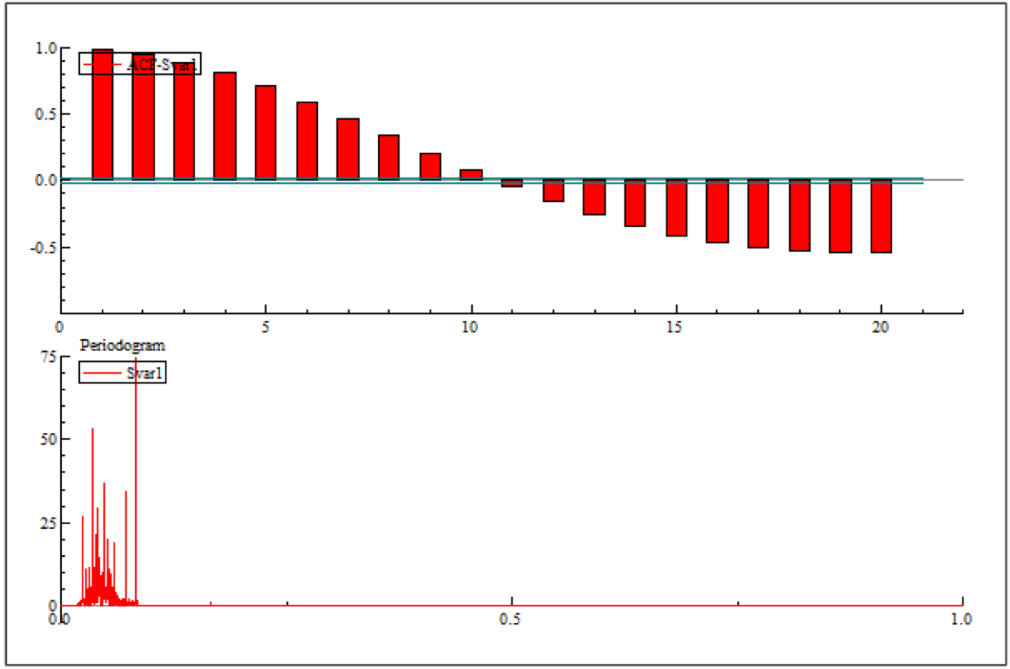

**Figure 8: Residuals ACF & Periodogram**

As shown in the figure (8), the periodogram of the $k$-factor GARMA residuals, the spectral density is unbounded at equidistant frequencies that indicates the presence of several seasonalities. Which requires the use of some seasonal long memory G-GARCH method to model such processes.

**Table 6: Estimation results of the G-GARCH model**

| Parameters | G-GARCH model estimation |
|---|---|
| $\hat{\psi}$ | 0.1341*** |
| $\hat{\beta}$ | 0.1652*** |
| $\hat{\gamma}$ | *0.0004**** |
| $\hat{d}_{v,1}$ | 0.2492*** |
| $\hat{d}_{v,2}$ | 0.1295*** |

| $\hat{\lambda}_{v,1}$ | 0.0178*** | (56h,17min $\approx$ 2days) |
|---|---|---|
| $\hat{\lambda}_{v,2}$ | 0.0416*** | (24h.038min $\approx$ 1 day) |

*Note: *** indicate rejection of the null hypothesis at the 1-percent level.*

The residuals from the $k$-factor GARMA are modelled using two different approaches in order to preserve the best estimation model. In the first approach, we adopt the G-GARCH model in order to estimate the seasonal long memory behaviour in the conditional variance. The estimation results reported in table 6. The spectral densities, represented by periodogram (figure 8), are unbounded at equidistant frequencies, which proves presence of several seasonality's. They show special peaks at frequencies $\hat{\lambda}_{v,1} = 0.0178$ (T=56h, 17min $\approx$ 2days) and $\hat{\lambda}_{v,2} = 0.0416$ (T=24h.038min $\approx$ 1 day), as shown in table 6, that corresponding to cycles with two day and daily periods, respectively.

**LLWNN estimation results**

In the second approach, residuals resulting from the $k$-factor GARMA modeling are considered as the input of the LLWNN, and shaped through the network in order to estimate the conditional variance. For the purpose of avoiding the possibility of coupling among different input and to accelerate convergence, all the inputs are normalized within a range of $[0, 1]$ using the most commonly used data smoothing method before applying it to the network.

$$y_{norm} = \frac{y_{org} - y_{min}}{y_{max} - y_{min}} \quad (30)$$

Where $y_{norm}$ is the normalized value, $y_{org}$ is the original value, $y_{min}$ and $y_{max}$ are the minimum and maximum values of the corresponding residuals data.

The datasets is divided into three successive parts as follows: (a) A sample of 200 observations to initialize the network training, (b) a training set and (c) a test set. The forecasting experiment is performed over the test set using an iterative forecasting scheme, the model are forecasting for 6, 12, 24, 48 and 72 hours ahead. Details of the datasets are given in the figure 9.

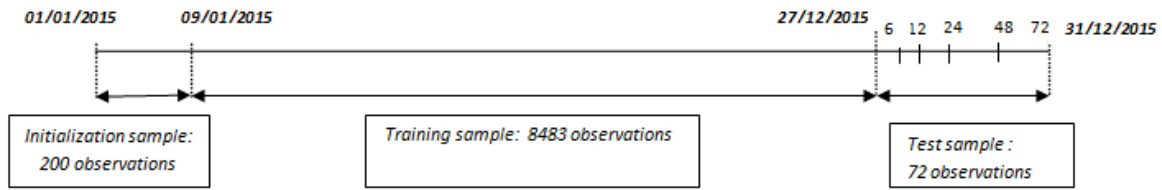

**Figure 9: Details of datasets.**

In order to find the best neural network architecture, at the beginning the parameters are randomly initialized. Thereafter, using two different algorithms: the Back Propagation algorithm (BP) and the Particle Swarm Optimization Algorithm (PSO), these parameters are updated in order to minimize the error between the output values and the real values during the training of the network. Table 7 and table 8 provide the summary of information related to the network architecture.

**Table 7: LLWNN based BP algorithm architecture**

| Number of hidden layer | 10 |
|---|---|
| Learning rate | 0.5 |
| Layer Activation function | Wavelet Function |
| Learning Algorithm | Back Propagation (BP) Learning Algorithm |

**Table 8: LLWNN based PSO algorithm architecture**

| Number of hidden layer | 10 |
|---|---|
| Learning rate | 0.5 |
| Layer Activation function | Wavelet Function |
| L earning Algorithm | Particle Swarm Optimization Algorithm |
| Number of particles | 20 |
| loc-best | 16 |

### 4.3. Forecasting results: A comparative Approach

This section is devoted to the evaluation of the estimated models in a multi-step-ahead forecasting task. Since forecasting is basically an out-of-sample problem, we prefer to apply out-of-sample criteria. Accordingly, five different periods (6 hours, 12 hours, one day, two days and tree days, ahead forecasting) were selected in aim to insure the quality and the robustness of modeling and forecasting results. In order to evaluate the forecasting accuracy, we apply five evaluation criteria, namely, the out-of-sample $R^2$ of Campbell and Thompson (2008), the mean absolute percentage error (MAPE), the logarithmic loss function(LL), the mean absolute error (MAE), the mean square error (MSE) and the root mean square error (RMSE) given respectively by

$$R^2 = 1 - \left( \frac{\sum_{t=t_1}^{N}(y_{t+h} - \hat{y}_{t,t+h})^2}{\sum_{t=t_1}^{N}(y_{t+h} - \bar{y}_{t,t+h})^2} \right), \quad (31)$$

$$MAPE = \frac{1}{N - t_1} \sum_{t=t_1}^{N} \left| \frac{(y_{t+h} - \hat{y}_{t,t+h})}{y_{t+h}} \right| * 100, \quad (32)$$

$$LL = \frac{1}{N - t_1} \sum_{t=t_1}^{N} \left( Log \left( \frac{\hat{y}_{t,t+h}}{y_{t+h}} \right) \right)^2. \quad (33)$$

$$MAE = \frac{1}{N - t_1} \sum_{t=t_1}^{N} \left| (y_{t+h} - \hat{y}_{t,t+h}) \right| \quad (34)$$

$$MSE = \frac{1}{N - t_1} \sum_{t=t_1}^{N} (y_{t+h} - \hat{y}_{t,t+h})^2 \quad (35)$$

$$RMSE = \left( \frac{1}{N - t_1} \sum_{t=t_1}^{N} (y_{t+h} - \hat{y}_{t,t+h})^2 \right)^{1/2} \quad (36)$$

Where $N$ is the number of observations, $N - t_1$ is the number of observations for predictive performance, $y_{t+h}$ is the log-return series through period $t + h$, $\hat{y}_{t,t+h}$ is the predictive log-return series of the predictive horizon $h$ at time $t$ and $\bar{y}_{t,t+h}$ is the historical average forecast.

**Table 9: Out of sample forecasts results**

| Models | Criterion | $h = 6$ | $h = 12$ | $h = 24$ | $h = 48$ | $h = 72$ |
|---|---|---|---|---|---|---|
| *k-factor* | $R^2$ | 0,9428 | 0.6040 | 0.9977 | 0.9528 | 0.9889 |
| *GARMA-* | MAPE | 0.9178% | 4.4437% | 0.5719% | 2.0084% | 0.9902% |
| *LLWNN based Back Propagation (BP) algorithm model* | LL | $1.122\times10^{-4}$ | 0.0021 | $5.734\times10^{-5}$ | $5.824\times10^{-4}$ | $1.556\times10^{-4}$ |
| | MAE | 0.0029 | 0.0147 | 0.0020 | 0.0080 | 0.0039 |
| | MSE | $1.143\times10^{-5}$ | $2.175\times10^{-4}$ | $6.953\times10^{-6}$ | $9.84\times10^{-5}$ | $2.386\times10^{-5}$ |
| | RMSE | 0.0034 | 0.0147 | 0.0026 | 0.0099 | 0.0049 |
| *The hybrid k-factor GARMA-LLWNN based Particle Swarm Optimization (PSO) algorithm model* | $R^2$ | 0.9989 | 0.9999 | 0.9994 | 0.9927 | 0.9971 |
| | MAPE | 0.1752% | 0.1031% | 0.2508% | 0.9897% | 0.9031% |
| | LL | $5.538\times10^{-6}$ | $1.423\times10^{-6}$ | $7.824\times10^{-6}$ | $1.735\times10^{-4}$ | $1.268\times10^{-4}$ |
| | MAE | $6.039\times10^{-4}$ | $3.090\times10^{-4}$ | $7.820\times10^{-4}$ | 0.0038 | 0.0032 |
| | MSE | $7.022\times10^{-7}$ | $1.255\times10^{-7}$ | $8.054\times10^{-7}$ | $2.781\times10^{-5}$ | $1.466\times10^{-5}$ |
| | RMSE | $8.379\times10^{-4}$ | $3.543\times10^{-4}$ | $8.974\times10^{-4}$ | 0.0053 | 0.0038 |
| *The k-factor GARMA-G-GARCH model* | $R^2$ | 0.9993 | 0.9998 | 0.9997 | 0.9976 | 0.9981 |
| | MAPE | 0.1564% | 0.0653% | 0.2376% | 0.9983% | 0.9568% |
| | LL | $4.328\times10^{-6}$ | $1.109\times10^{-6}$ | $6.567\times10^{-6}$ | $1.437\times10^{-4}$ | $1.172\times10^{-4}$ |
| | MAE | $5.748\times10^{-4}$ | $2.563\times10^{-4}$ | $6.112\times10^{-4}$ | $1.352\times10^{-4}$ | $1.172\times10^{-4}$ |
| | MSE | $6.557\times10^{-7}$ | $1.105\times10^{-7}$ | $7.372\times10^{-7}$ | $1.981\times10^{-5}$ | $1.238\times10^{-5}$ |
| | RMSE | $6.785\times10^{-4}$ | $2.768\times10^{-4}$ | $7.647\times10^{-4}$ | $1.981\times10^{-5}$ | $1.986\times10^{-4}$ |

In order to evaluate the prediction performance of the proposed hybrid methodology, this paper has taken into account two approaches: the hybrid $k$-factor GARMA-LLWNN model trained using two different algorithms (the Backpropagation algorithm and the Particle Swarm Optimization algorithm), and the $k$-factor GARMA-G-GARCH model. And five time horizons; 6 hours, 12 hours, one day, 2 days and 3 days ahead forecasting, using the R², the MAPE, the LL, the MAE, the MSE and the RMSE out of sample criteria, the forecast evaluation results are reported in Table 9.

It shows that the $k$-factor GARMA-LLWNN model based PSO algorithm outperforms the $k$-factor GARMA-LLWNN model based BP algorithm, hence, the PSO algorithm is more efficient than the BP algorithm. As shown in Figures 11, 12, 13, 14, 15, 17, 18, 19, 20 and 21 the predictions of the $k$-factor GARMA-LLWNN model based PSO algorithm for all the five horizons are very close to the real values.

In addition, the $k$-factor GARMA-G-GARCH model outperforms the hybrid model in terms of prediction accuracy. Indeed, the $k$-factor GARMA-G-GARCH model prediction errors are the smallest for all evaluation criteria and for all forecast time horizons.

This can be explained by the fact that the $k$-factor GARMA-G-GARCH model takes into account the seasonal long-memory in both the conditional mean and the conditional variance, making this model a robust tool that can deal with the features of the electricity prices and thus, provide the best forecasting results. On the other hand, despite its capacity as a nonlinear, nonparametric model, and its particularity by having a wavelet activation function and local linearity, the LLWNN model is unable to detect, model and predict the seasonality and the long memory dependence in the data. Since, when it is compared with the G-GRACH model, this last one provides prediction that is more accurate. This is explained by the ability of the G-GARCH model in modeling the seasonal long memory in the conditional variance. This also proves the importance of taking into account the seasonal long memory behavior to enhance forecasting accuracy [Nowotarski and Weron 2016].

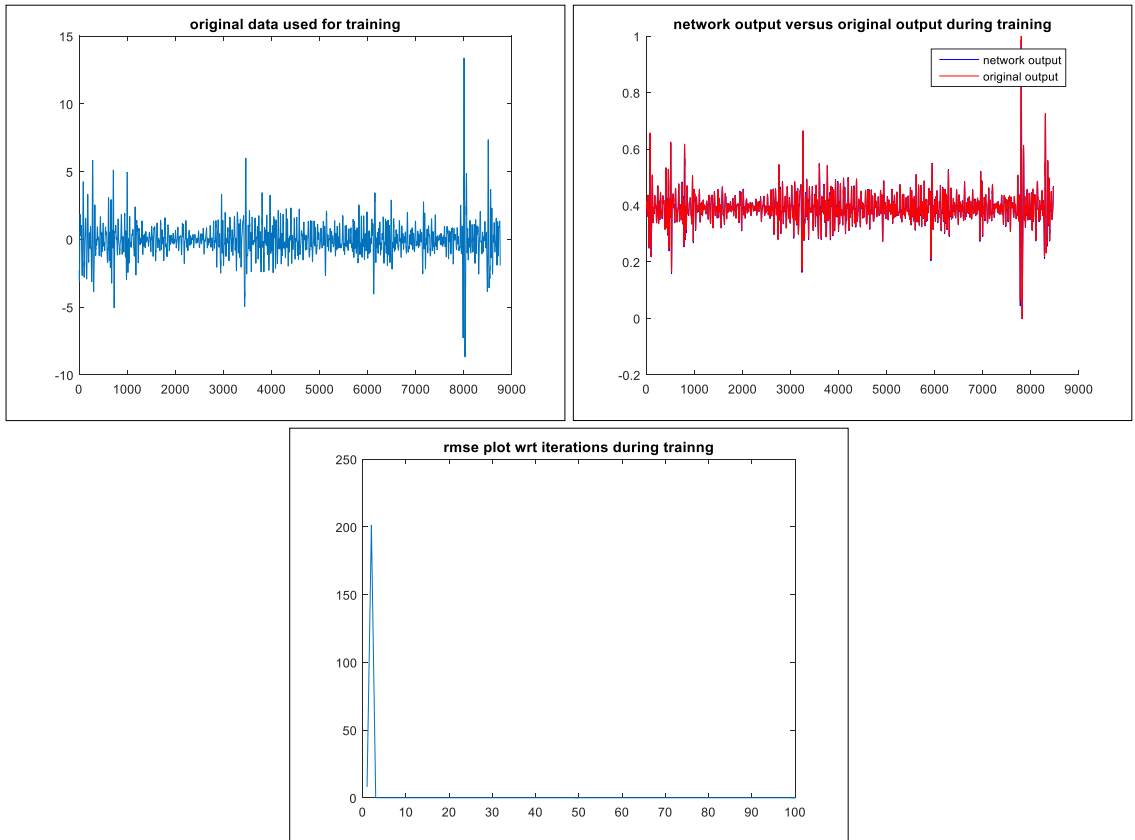

**Figure 10: LLWNN Training with BP algorithm results (residuals of GARMA modeling)**

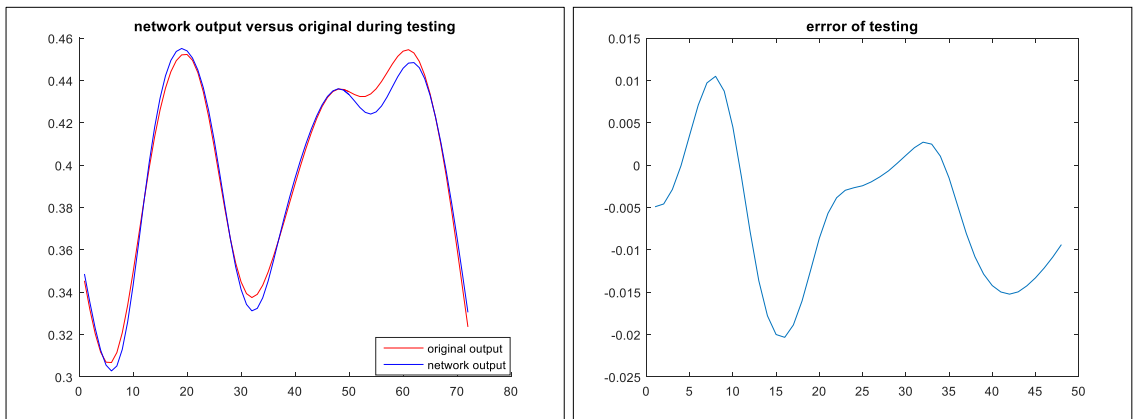

**Figure 11: Tree days (72 hours) ahead prediction during testing (residuals of GARMA modeling)**

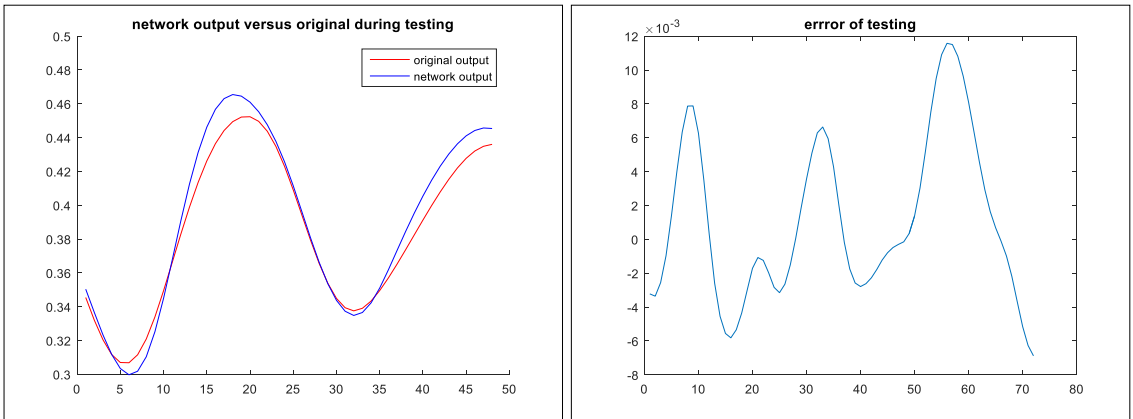

**Figure 12: Two days (48 hours) ahead prediction during testing (residuals of GARMA modeling)**

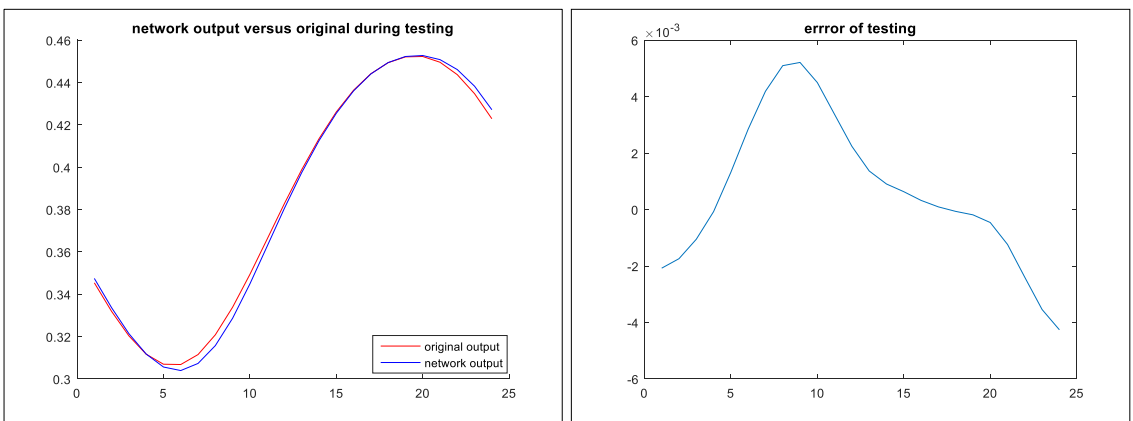

**Figure 13: One days (24 hours) ahead prediction during testing (residuals of GARMA modeling)**

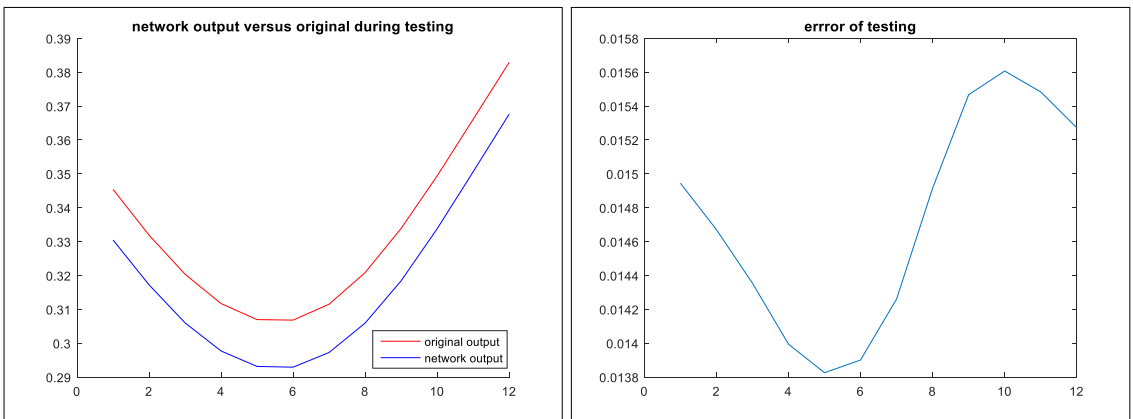

**Figure 14: Semi daily (12 hours) prediction during testing (residuals of GARMA modeling)**

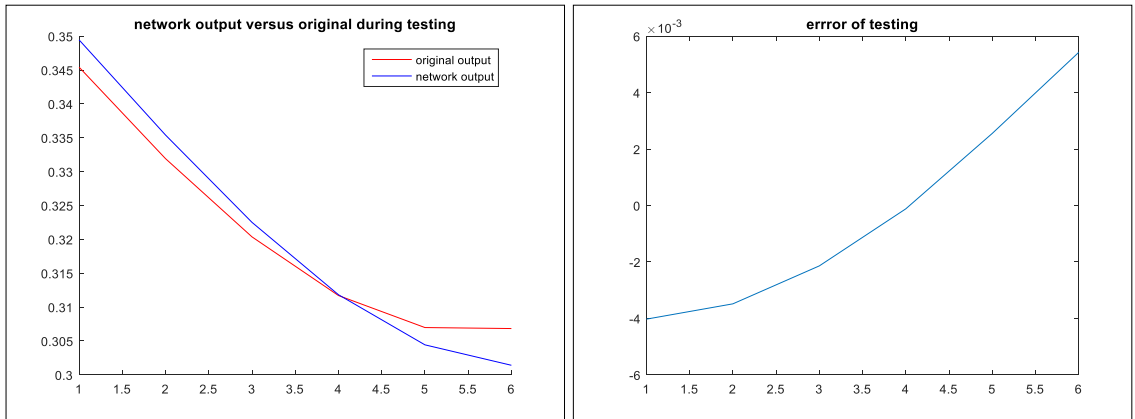

**Figure 15: 6 hours ahead prediction during testing (residuals of GARMA modeling)**

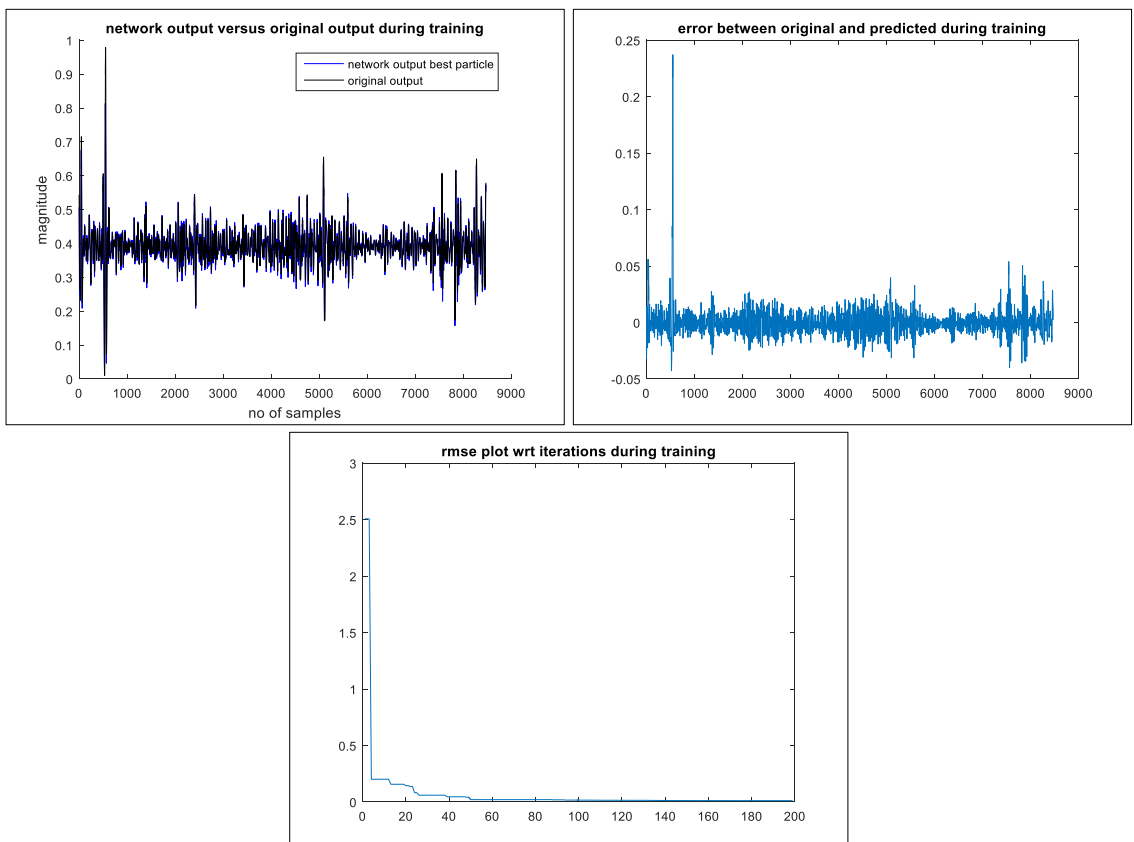

**Figure 16: LLWNN Training with PSO algorithm results (residuals of k-factor GARMA modeling)**

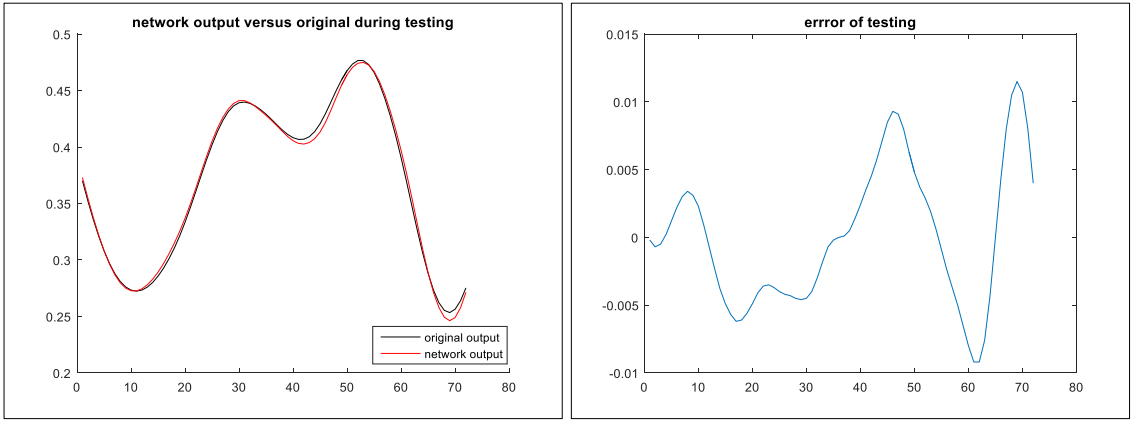

**Figure 17: Three days (72h) ahead prediction during testing (residuals of k-factor GARMA modeling)**

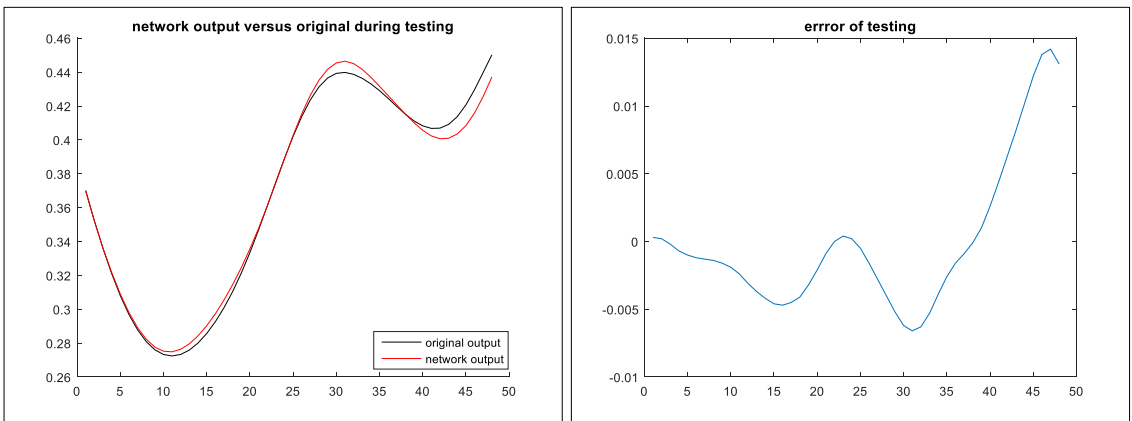

**Figure 18: Two days (48h) ahead prediction during testing (residuals of k-factor GARMA modeling)**

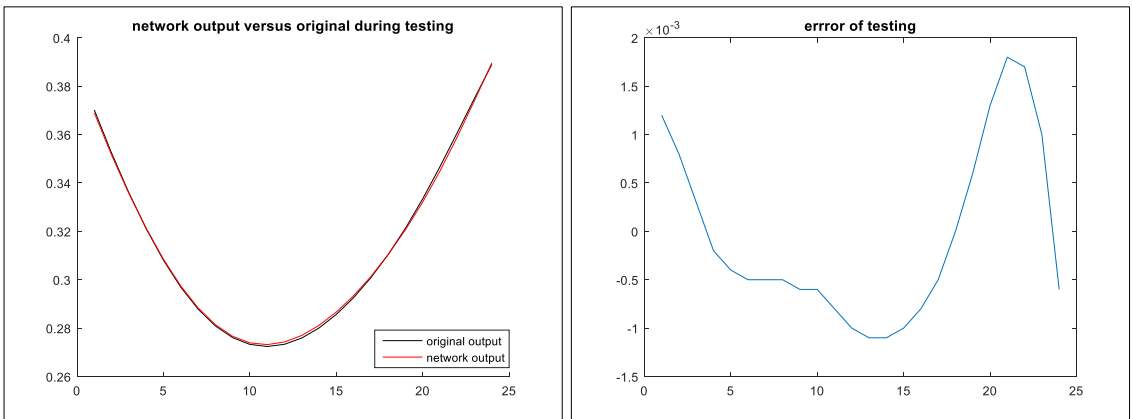

**Figure 19: One day ahead prediction during testing (residuals of k-factor GARMA modeling)**

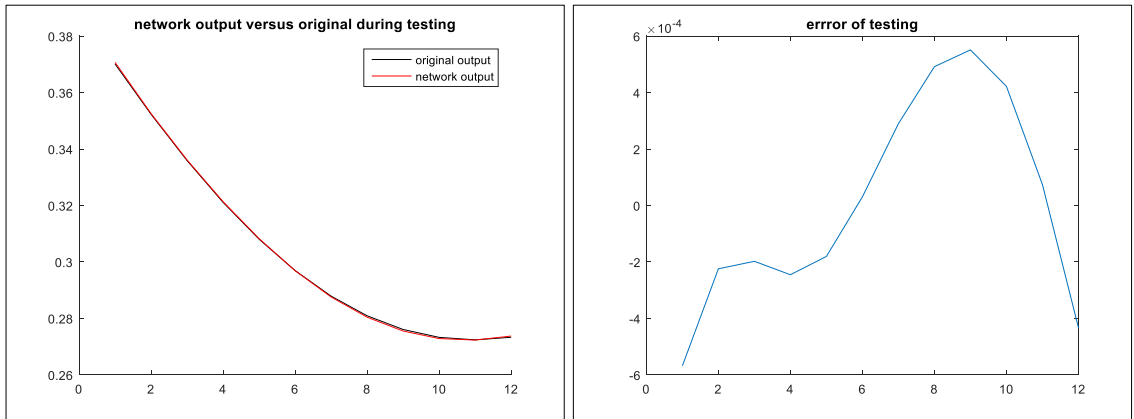

**Figure 20: 12 hours ahead prediction during testing (residuals of k-factor GARMA modeling)**

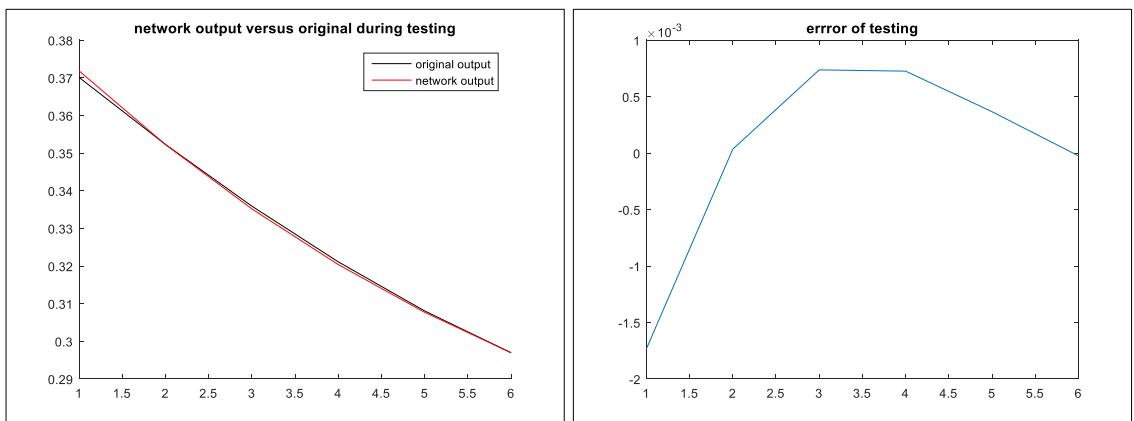

**Figure 21: 6 hours prediction during testing (residuals of k-factor GARMA modeling)**

## 5 Conclusions

In power markets, price analysis has become an important topic for all its participants. Background information about the electricity price is crucial for risk management. It may represent an advantage for a market player facing competition. Forecasting electricity prices at different periods is valuable for all industry stakeholders for cash flow analysis, financial procurement, capital budgeting, regulatory rule making, and integrated resource planning.

However, the behavior of electricity prices differs from that of other commodity markets. The most obvious of these differences is that electricity is a non-storable merchandize.

Moreover, electricity prices show some particular characteristics such as high frequency, non-stationary behavior, multiple seasonality, high volatility, hard nonlinear behavior and long memory, which may affect the prices dramatically. In this respect, there is no other market like it [Weron 2006], thus, we cannot rely on models developed for financial markets or other commodity markets. In this framework, due to the complexity of the electricity market, the electricity price forecasting has been the most challenging task. This has also motivated the researchers to develop efficient and intelligent methods to forecast the prices so that all stakeholders in the market can benefit out of it.

In this framework, this paper focuses on resolving the issues of modeling and forecasting the features of the electricity prices, notably the existing of the seasonal long memory behavior in both the conditional mean and the conditional variance.

For this purpose, we focus on the modeling of the conditional mean so we adopt a generalized fractional model with $k$-factor of Gegenbauer ($k$-factor GARMA). Thereafter, in order to model and predict the conditional variance we adopt two different approaches; firstly, the local linear wavelet neural network (LLWNN) is adopted to model and predict the conditional variance (applied to the residual of $k$-factor GARMA model). And we adopt two different learning algorithms; the Back Propagation and the Particle Swarm Optimization, in order to preserve the algorithm that optimize the parameters of the network. Secondly, the G-GARCH model is applied to the residual of the $k$-factor GARMA, so we estimate a $k$-factor GARMA-G-GARCH model and we use an estimation approach based on the discrete wavelet transform (DWPT). More precisely, we estimated a $k$-factor GARMA-G-GARCH model using the wavelet based approximate maximum likelihood estimator developed by Whitcher (2004). The $R^2$ of Campbell and Thompson (2008), the mean absolute percentage error (MAPE), the logarithmic loss function (LL), the mean absolute error (MAE), the mean square error (MSE) and the root mean square error (RMSE) are used as a performance index to evaluate the prediction capability of each model.

The predictive performance of $k$-factor GARMA-G-GARCH model provides evidence of the power compared to the hybrid parametric and non-parametric $k$-factor GARMA-LLWNN model. Therefore, this model leads to an improved performance. It can be an effective way in the forecasting task, especially when higher forecasting accuracy is needed. Obtained results are very interesting in the meaning that it was always difficult to accomplish such precision when forecasting electricity spot prices. This highlights the importance of the $k$-factor GARMA-G-GARCH methodology as a robust forecasting method.

**References**

Andersen, T.G., and T. Bollerslev. 1997. "Intraday periodicity and volatility persistence in financial markets". *J. Emp. Finance 4: 115-158.*

Armano, G., M. Marchesi, and A. Murru. 2005. "A hybrid genetic-neural architecture for stock indexes forecasting". *Information Sciences 170: 3-33.*

Baillie, R.T., T. Bollerslev, and H.O. Mikkelsen. 1996. "Fractionally integrated generalized autoregressive conditional heteroscedasticity". *J. Econ 74 (1): 3- 30.*

Bashir, Z., and M.E. El-Hawary. 2000. "Short term load forecasting by using wavelet neural networks, in: Electrical and Computer Engineering." *IEEE 163-166.*

Benaouda, D., G. Murtagh, J.L. Starck, and O. Renaud. 2006. "Wavelet-based nonlinear multiscale decomposition model for electricity load forecasting". *Neurocomputing, 70, 139-154.*

Beran, J. 1999. "SEMIFAR Models-a Semiparametric Framework for Modelling Trends, Long-Range Dependence and Nonstationarity". *Center of Finance and Econometrics, University of Konstanz.*

Bernard, C., Mallat, S., and J.J. Slotine .1998. "Wavelet interpolation networks". *ESANN: 49-52.*

Bollerslev, T. 1986. "Generalized autoregressive conditional heteroscedasticity". *Journal of Econometrics 31: 307-327.*

Bollerslev, T., and R. Hodrick. 1992. "Financial market efficiency tests". *The Handbook of Applied Econometrics I. Macroeconomics, North-Holland, Amsterdam.*

Bollerslev, T., and H.O. Mikkelsen. 19996. "Modeling and pricing long memory in stock market volatility". *J. Econometrics 73: 151-184.*

Bordignon, S., M. Caporin, and F. Lisi. 2010. "Periodic long memory GARCH models" *Econom Rev 28:60-82.*

Bordignon, S., M. Caporin, and F. Lisi. 2007. "Generalised long-memory GARCH models for intra-daily volatility". *Computational Statistics & Data Analysis 51: 5900-5912.*

Boubaker, H. 2015. "Wavelet Estimation of Gegenbauer Processes: Simulation and Empirical Application". *Computational Economics 46: 551-574.*

Boubaker; H., and M. Boutahar. 2011. "A wavelet-based approach for modelling exchange rates". *Stat Methods Appl 20: 201-220.*

Boubaker, H., and N. Sghaier. 2015. "Semi parametric generalized long-memory modeling of some MENA stock market returns: A wavelet approach." *Economic Modeling 50: 254-*

*265.*

Breidt, F.J., N. Crato, and P. de Lima. 1998. "The detection and estimation of long memory in stochastic volatility". *J. Econometrics 83: 325-348.*

Campbell, J.Y., and S.B. Thompson. 2008. "Predicting the equity premium out of sample: can anything beat the historical average." *Rev. Finance. Stud. 21: 1509-1531.*

Cao, L., Y. Hong, H. Fang, and G. He. 1995. "Predicting chaotic time series with wavelet Networks." *Physica D 85: 225-238.*

Caporale, G.M., J. Cuñado, and L.A. Gil-Alana. 2012. "Modelling long-run trends and cycles in financial time series data." *J. Time Ser. Anal. 34 (2): 405-421.*

Caporale, G.M., L.A. Gil-Alana. 2014. "Long-run and cyclical dynamics in the US stock market." *J. Forecast. 33 (2): 147-161.*

Caporin, M., and F. Lisi. 2010. "Misspecification tests for periodic long memory GARCH models." *Statistical Methods & Applications 19: 47-62.*

Chakravarty, S., M. Nayak, and M. Bisoi. 2012. "Particle Swarm Optimization Based Local Linear Wavelet Neural Network for Forecasting Electricity Prices, Energy, Automation, and Signal (ICEAS)." *Energy, Automation, and Signal (ICEAS), IEEE: 1-6.*

Chen, Y., J. Dong, B. Yang, and Y. Zhang, 2004. "A local linear wavelet neural network." *Intelligent Control and Automation, IEEE: 1954-1957.*

Chen, Y., B. Yang, and J. Dong. 2006. "Time-series prediction using a local linear wavelet neural network." *Neurocomputing 69 (2006) 449-465.*

Cheung, Y.W. 1993. "Long memory in foreign-exchanges rates." *J Bus Econ Stat 11: 93-101.*

Clewlow, L., C. Strickland. 2000. "Energy derivatives pricing and risk management." *Lacima Publications, Lonfon.*

Contreras, J., R. Espinola, F.J. Nogales, and A.J. Conejo. 2003. "ARIMA models to predict next day electricity prices." *IEEE Transactions on Power systems 18(3): 1014-1020.*

Cristea, P., R. Tuduce, and A. Cristea. 2000. "Time series prediction with wavelet neural networks." *Neural network applications in electrical engineering.*

Diongue, A.K., G. Dominique, and V. Bertrand. 2004. "A k-factor GIGARCH process: estimation and application on electricity market spot prices." *Probabilistic Methods Applied to Power Systems: 12-16.*

Diongue, A.K., and Guégan. D. 2008. "Estimation of k-factor GIGARCH process: a Monte Carlo study." *ISSN : 1955-611X.*

Diongue, A.K., D. Guégan, and B. Vignal. 2009. "Forecasting electricity spot market prices with a k-factor GIGARCH process" *Applied Energy 86: 505-510.*

Engle, R.F., 1982. "Autoregressive conditional heteroscedasticity with estimates of the variance of United Kingdom inflation." *Econometrica 50: 987-1008.*

Escribano, A., J.I. Peña, and P. Villaplana. 2011. "Modelling electricity prices: international evidence." *Oxford bulletin of economics and statistics 73 : 622-650.*

Ferrara, L., and D. Guegan. 2001. "Forecasting with k-factor Gegenbauer Processes: Theory and Applications." *Journal of Forecasting, Wiley 20 (8): 581-601.*

Ferrara, L., and D. Guegan. 2007. "Fractional seasonality: Models and Application to Economic Activity in the Euro Area". *Eurostat Pulications-Luxembourg. Eurostat Pulications: 137-153.*

Gao, R., and H. I. Tsoukalas. 2001. "Neural-wavelet methodology for load forecasting." *Journal of Intelligent & Robotic Systems 31: 149-157.*

Gencay, R., M.M. Dacorogna, U.A. Müller, O.V. Pictet, and R.B. Olsen. 2001. "An Introduction to High Frequency Finance." *Academic Press, London.*

Geweke, J., and S. Porter-Hudak. 1983. "The Estimation And Application Of Long Memory Time Series Models." *Journal of Time Series Analysis 4: 221–238.*

Giraitis, L., P. Kokoszka, and R. Leipus. 2000. "Stationary ARCH models: dependence structure and Central Limit Theorem." *Econometric Theory 16: 3-22.*

Granger, C.W., R. Joyeux. 1981. "An introduction to long-memory time series models and fractional differencing." *Journal of time series analysis 1, 15–29.*

Gray, H.L., N.F. Zhang, and W.A. Woodward. 1989. "On generalized fractional processes". *Journal of Time Series Analysis 10 (3): 233-257.*

Guégan, D. 2000. "A New Model: The k-factor GIGARCH Process." *Journal of Signal Processing 4: 265-271.*

Ham F.H. and I. Koslanic. 2001. "Principles of Neurocompuling for Science and Engineering." *McGraw-Hill Higher Education.*

Hosking, J.R.M. 1981. "Fractional differencing." *Biometrika 68 (1): 165-176.*

Kazakeviéius. V., and R. Leipus. 2002. "On stationarity in the ARCH($\infty$) model." *Econometric Theory 18: 1-16.*

Khashei, M. and M. Bijari. 2010. "An artificial neural network model for time series forecasting." *Expert Syst. Appl. 37: 479-489.*

Khashei, M., and M. Bijari. 2011. "A novel hybridization of artificial neural networks and ARIMA models for time series forecasting." *Applied Soft Computing 11: 2664-2675.*

Knittel, C., and M.R. Roberts. 2005. "An empirical examination of restructured electricity prices." *Energy Economics 27(5): 791-817.*

Koopman, S.J., M. Ooms, and M.A Carnero. 2007. "Periodic seasonal Reg ARFIMA-GARCH models of daily electricity spot prices." *Journal of the American Statistical Association 102(477): 16-27.*

Lina, L., Y. Lib, and A. Sadek. 2013. "A k Nearest Neighbor based Local Linear Wavelet Neural Network Model for On-line Short-term Traffic Volume Prediction." *Procedia - Social and Behavioral Sciences 96: 2066-2077.*

Mallat, S. 1999. "A wavelet tour of signal processing". *Academic press.*

Mallat, S., and Zhang. 1993. "Matching pursuits with time-frequency dictionaries." *IEEE Transactions on Signal Processing 41: 3397-3415.*

Mohapatra. P., M. Anirudh, and T.K. Patra. 2013. "Forex Forecasting: A Comparative Study of LLWNN and NeuroFuzzy Hybrid Model." *International Journal of Computer Applications 66: No.18.*

Nowotarski, J., and R. Weron. 2016. "On the importance of the long-term seasonal component in day-ahead electricity price forecasting." *Energy Economics 57: 228-235.*

Pany, P.K. 2011. **"**Short-Term Load Forecasting using PSO Based Local Linear Wavelet Neural Network." *International Journal of Instrumentation 1: Issue-2.*

Pany, P.K., and S.P. Ghoshal. 2013. "Day-ahead Electricity Price Forecasting Using PSO-Based LLWNN Model." *International Journal of Energy Engineering (IJEE) 3: 99-106.*

Pany, P.K., and S.P. Ghoshal. 2015. "Dynamic electricity price forecasting using local linear wavelet neural network." *Neural Comput & Applic.*

Pati, Y.C., and P.S. Krishnaprasad. 1993. "Analysis and synthesis of feedforward neural networks using discrete affine wavelet transforms." *IEEE Transactions on Neural Networks, 4(1): 73-85.*

Percival, D.B., and A.T. Walden. 2000. "Wavelet methods for time series analysis." *Cambridge University Press, Cambridge.*

Robinson, P.M. 1995. "Log-Periodogram Regression of Time Series with Long-Range Dependance." *Annals of statistics 23: 1048-1072.*

Saâdaoui, F., N. Chaâben, and S. Benammou. 2012. "Modelling power spot prices in deregulated European energy markets: a dual long memory approach." *Global Business and Economics Review 14: No. 4.*

Sharkey, A.J., 2002. Types of multinet system, in: International Workshop on Multiple Classifier Systems. Springer, pp. 108-117.

Soares, L.J., and L.R. Souza. 2006. "Forecasting electricity demand using generalized long memory." International Journal of Forecasting 22, 17-28.

Szkuta. B., L. Sanabria, and T. Dillon. 1999. "Electricity price short-term forecasting using artificial neural networks." *IEEE Trans Power Syst 14: 851-857.*

Tan. Z., J. Zhang, J. Wang, and J. Xu. 2010. "Day-ahead electricity price forecasting using wavelet transform combined with ARIMA and GARCH models."*Appl Energy 87:3606-3610.*

Taskaya. T., and M.C. Casey. 2005. " A comparative study of autoregressive neural network hybrids." *Neural Networks 18: 781-789.*

Tong; H., K.S. Lim. 1980. "Threshold autoregressive, limit cycles and cyclical data." *Journal of the Royal Statistical Society Series 42 (3): 245-292.*

Tseng, F.M., H.C. Yu, and G.H. Tzeng. 2002. "Combining neural network model with seasonal time series ARIMA model." *Technological Forecasting & Social Change 69: 71-87.*

Ulugammai, M., P. Venkatesh, P.S. Kannan, and N.P. Padhy. 2007. "Application of bacterial foraging technique trained artificial and wavelet neural networks in load forecasting." *Neurocomputing 70: 2659-2667.*

Valenzuela, O., I. Rojas, F. Rojas, H. Pomares, L. Herrera, A. Guillen, L. Marquez, and M. Pasadas. 2008. "Hybridization of intelligent techniques and ARIMA models for time series prediction." *Fuzzy Sets Syst. 159: 821-845.*

Wang, A., and B. Ramsay. 1998. "A neural network based estimator for electricity spot-pricing with particular reference to weekend and public holidays." *Neurocomputing 23: 47-57.*

Weron. R., I. Simonsen, and P. Wilman. 2004. "Modelling highly volatile and seasonal markets: evidence from the NordPool electricity market." *The Application of Econophysics, Springer, Tokyo 182-191.*

Weron, R. 2006. "Modeling and Forecasting Electricity Loads and Prices: A Statistical Approach." *The Wiley Finance Series (Book 396), Wiley, Chichester.*

Whitcher, B. 2004. "Wavelet-based estimation for seasonal long-memory processes." *Technometrics 46.*

Woodward, W.A., Q.C. Cheng, and H.L. Gray. 1998. "A k-factor GARMA long-memory model." *J Time Series Analysis 19: 485-504.*

Yao S.J., Y.H. Song, L.Z. Zhang, and X.Y. Cheng. 2000. "Wavelet transform and neural networks for short-term electrical load forecasting." *Energy Conversion & Management 41: 1975-1988.*

Yu, L., S. Wang, and K.K. Lai. 2005. A novel nonlinear ensemble-forecasting model incorporating GLAR and ANN for foreign exchange rates." *Computers and Operations Research 32: 2523-2541.*

Zaffaroni, P., 2004. "Stationarity and memory in ARCH($\infty$) models." *Econometric Theory 20: 147-160.*

Zhang, Q. 1997. "Using wavelet network in nonparametric estimation." IEEE Transactions on Neural Networks 8(2): 227-236.

Zhang, Q., and A. Benveniste. 1992. "Wavelet networks." *IEEE Transactions on Neural Networks, 3(6): 889-898.*

Zhang, G.P. 2003. "Time series forecasting using a hybrid ARIMA and neural network model." *Neurocomputing 50: 159- 175.*